\documentclass[11pt]{article}
\usepackage{amssymb}
\usepackage{amsmath}
\usepackage{graphics}
\usepackage{epsfig}
\usepackage{a4wide}
\usepackage{cite}
\usepackage{multirow}
\usepackage{color}
\usepackage{mathrsfs}
\usepackage{slashed}

\DeclareGraphicsExtensions{.eps}
\begin{document}

\title{$H_5^{\pm\pm}h^0$ production via vector-boson fusion in Georgi-Machacek model at hadron colliders}
\author{
Qiang Yang,$^{1,2}$ Ren-You Zhang,$^{1,2}$\footnote{Corresponding author. zhangry@ustc.edu.cn}~ Wen-Gan Ma,$^{1,2}$ Yi Jiang,$^{1,2}$ Xiao-Zhou Li,$^{1,2}$\\ and Hao Sun$^{3}$ \\ \\
{\small $^1$ State Key Laboratory of Particle Detection and Electronics,} \\
{\small University of Science and Technology of China, Hefei 230026, Anhui, People's Republic of China} \\
{\small $^2$ Department of Modern Physics, University of Science and Technology of China,}  \\
{\small Hefei 230026, Anhui, People's Republic of China} \\
{\small $^3$ Institute of Theoretical Physics, School of Physics, Dalian University of Technology,}  \\
{\small Dalian 116024, Liaoning, People's Republic of China} }

\date{}
\maketitle
\vskip 10mm

\begin{abstract}
The Georgi-Machacek (GM) model is a distinctive TeV-scale extension of the Standard Model (SM) due to the introduction of two (one real and one complex) scalar triplets to the Higgs sector. It predicts the existence of doubly charged Higgs bosons $H_5^{\pm\pm}$ and the vacuum expectation value of Higgs triplets $v_\Delta$ can reach a few tens of GeV. In this paper, we perform a parameter scan of the GM model within the H5plane benchmark scenario and investigate in detail the single production of a doubly charged Higgs boson in association with a SM-like Higgs boson via vector-boson fusion at the $14~ {\rm TeV}$ LHC and $70~ {\rm TeV}$ Super Proton-Proton Collider. Both integrated cross section and differential distributions with respect to some kinematic variables for $pp \rightarrow W^{\pm}W^{\pm} \rightarrow H_5^{\pm\pm} h^0 + 2\, {\rm jets}$ are provided up to the QCD next-to-leading order. In the signal-background analysis we employ the {\sc madspin} method to take into account the spin correlation and finite width effects of the intermediate Higgs and $W$ bosons and present some kinematic distributions of final leptons. The numerical results show that the SM background can be suppressed remarkably and the $H_5^{\pm\pm} h^0$ vector-boson fusion signal can be directly detected at future high-luminosity, high-energy hadron colliders by imposing a proper cut on the transverse mass $M_{T, \ell_1 \ell_2}$.
\end{abstract}


\vfill \eject
\baselineskip=0.32in
\makeatletter      
\@addtoreset{equation}{section}
\makeatother       
\vskip 5mm
\renewcommand{\theequation}{\arabic{section}.\arabic{equation}}
\renewcommand{\thesection}{\Roman{section}.}
\newcommand{\nb}{\nonumber}
\vskip 5mm
\section{INTRODUCTION}
\par
The $125~ {\rm GeV}$ Higgs boson has been discovered at CERN Large Hadron Collider (LHC)\cite{SM-higgs1,SM-higgs2}. The structure of the Higgs sector plays a crucial role for understanding the electroweak symmetry breaking and the mass origin of fundamental particles. In addition to the precise measurement of gauge, Yukawa, and self-couplings of the $125~ {\rm GeV}$ Higgs boson, the search for the exotic Higgs bosons predicted by some new physics models is an important task of the LHC and future hadron and lepton colliders. These new physics models may give a hint to solve the problems of the Standard Model (SM), such as the gauge hierarchy, the candidate for dark matter, the neutrino oscillation, and mass origin of neutrinos, etc.

\par
Among all the extensions of the SM, the Georgi-Machacek (GM) model\cite{GM-model1,GM-model2} is interesting due to the existence of Higgs triplets. The Higgs sector of the GM model consists of the following three $SU(2)_L$ scalar multiplets,
\begin{eqnarray}
&&\chi \equiv (\chi^{++},\chi^{+},\chi^{0}) :~~~~~ {\rm complex~ triplet},~~~~ Y=2, \nonumber \\
&&\,\xi \equiv (\xi^+,\xi^0,\xi^-) :~~~~~~~~ {\rm real~ triplet},~~~~~~~~~~ Y=0, \nonumber \\
&&\phi \equiv (\phi^+,\phi^0) :~~~~~~~~~~~~{\rm complex~ doublet},~~\, Y=1,~~~~~~
\end{eqnarray}
where $Y$ is the hypercharge and $(\phi^+,\phi^0)$ is the SM Higgs doublet. The electroweak parameter $\rho \equiv M_W^2/(M_Z^2\cos^2\theta_W)$ is kept to be unitary at tree level due to the custodial symmetry\cite{GM-model2} and the vacuum expectation value of the triplets $v_\Delta$ can reach a few tens of GeV. The GM model predicts many more scalars than the SM, including one Higgs fiveplet $H_5 = (H_5^{++}, H_5^{+}, H_5^{0}, H_5^{-}, H_5^{--})$, one Higgs triplet $H_3 = (H_3^+, H_3^0, H_3^-)$, and two Higgs singlets $H^0$ and $h^0$. The weak gauge couplings of fiveplet Higgs bosons are proportional to $v_\Delta$ at tree level in the GM model. However, these couplings are absent in the doublet extensions (e.g., the two Higgs doublet model) and are stringently constrained in some other triplet models (e.g., the left-right symmetric model). Thus, a distinct signal of the GM model is the production of a doubly charged Higgs boson with subsequent decay to two same-sign weak gauge bosons at hadron colliders\cite{search-higgs-at-lhc}.

\par
The phenomenology of the exotic Higgs bosons in the GM model at future electron-positron colliders was investigated in Ref.\cite{phenomenology-at-ee}. In Ref.\cite{test-custodial} the custodial symmetry was discussed in the fiveplet Higgs production via vector-boson fusion (VBF) at the LHC. The fiveplet Higgs VBF production, pair production, and associated production with a vector boson at the LHC in the GM model at the QCD next-to-leading order (NLO) including parton shower matching have been studied in Ref.\cite{automatic-nlo} by employing {\sc MadGraph5\_aMC@NLO}\cite{madgraph}. The inclusive cross section for single Higgs production via VBF at the LHC in the GM model has been calculated up to the next-to-next-to-leading-order (NNLO) accuracy in QCD by using the so-called structure function method in Ref.\cite{gm-sf-nnlo}.

\par
Compared to the single production, the pair production of Higgs boson can be used to test the strengths of Higgs self-couplings, which are extremely significant for understanding the electroweak symmetry breaking. In this paper we focus on the production of doubly charged Higgs boson $H_5^{\pm\pm}$ in association with a SM-like Higgs boson $h^0$ via VBF at hadron colliders. The VBF mechanism of $H_5^{\pm\pm} h^0$-associated production provides a clean experimental signature of two centrally produced Higgs bosons with two hard jets in the forward-backward rapidity region \cite{vbf}. The presence of the SM-like Higgs boson in the final state is a unique feature in tagging the $pp \rightarrow W^{\pm}W^{\pm} \rightarrow H_5^{\pm\pm} h^0 + 2\, {\rm jets}$ process, and the SM background to this VBF-associated production channel is reduced apparently in comparison with the VBF single production of the doubly charged Higgs boson.

\par
The rest of this paper is organized as follows. In Sec. II, we briefly review the Georgi-Machacek model. In Sec. III, we provide the numerical results of both integrated and differential cross sections for the associated production of $H_5^{\pm\pm} h^0$ via VBF in the H5plane benchmark scenario of the GM model at the QCD NLO and discuss the signal and the corresponding SM background of this process. Finally, a short summary is given in Sec. IV.

\section{GEORGI-MACHACEK MODEL}
The Higgs sector of the GM model consists of three isospin multiplets: a complex doublet $\phi$ with $Y=1$, which is identified as the SM Higgs doublet, a real triplet $\xi$ with $Y=0$, and a complex triplet $\chi$ with $Y=2$. All these Higgs fields can be written in the form of a bidoublet $\Phi$ and a bitriplet $\Delta$ under the global $SU(2)_L\times SU(2)_R$ symmetry,
\begin{eqnarray}
\Phi = \left(
       \begin{array}{cc}
       \phi^{0\ast} & \phi^+ \\
       -\phi^{-} & \phi^0
       \end{array}
       \right),
       ~~~~~~~~
\Delta = \left(
         \begin{array}{ccc}
         \chi^{0\ast} & \xi^+ & \chi^{++} \\
         -\chi^{-} & \xi^0 & \chi^+ \\
         \chi^{--} & -\xi^{-} & \chi^0
         \end{array}\right),
\end{eqnarray}
where $\phi^{-}=\phi^{+\ast}$, $\chi^{-}=\chi^{+\ast}$, and $\chi^{--}=\chi^{++\ast}$. The most general $SU(2)_L \times SU(2)_R \times U(1)_Y$ gauge invariant Higgs potential, which can ensure that $\rho = 1$ at tree level, is given by
\begin{align}
V_H ~=~ & \frac{\mu_2^2}{2}{\rm Tr}(\Phi^{\dagger} \Phi) + \frac{\mu_3^2}{2}{\rm Tr}(\Delta^{\dagger} \Delta)
	    + \lambda_1 [{\rm Tr}(\Phi^{\dagger} \Phi)]^2 + \lambda_2 {\rm Tr}(\Phi^{\dagger}\Phi){\rm Tr}(\Delta^{\dagger}\Delta) \nonumber\\
	  & + \lambda_3 {\rm Tr}[(\Delta ^{\dagger} \Delta)^2]
	    + \lambda_4 [{\rm Tr}(\Delta^{\dagger}\Delta)]^2
        - \lambda_5 {\rm Tr}(\Phi^{\dagger}\frac{\tau ^a}{2}\Phi \frac{\tau ^b}{2}) {\rm Tr}(\Delta^{\dagger}t^a \Delta t^b) \nonumber\\
      & - M_1 {\rm Tr}(\Phi^{\dagger}\frac{\tau^a}{2}\Phi\frac{\tau^b}{2})(P^{\dagger}\Delta P)^{ab}
        - M_2 {\rm Tr}(\Delta^{\dagger} t^a \Delta t^b)(P^{\dagger}\Delta P)^{ab},
\end{align}
where $\tau^a~ (a = 1, 2, 3)$ are the Pauli matrices,
\begin{eqnarray}
      t^1= \frac{1}{\sqrt{2}}\left(
            \begin{array}{ccc}
                  0 & 1 & 0 \\
                  1 & 0 & 1 \\
                  0 & 1 & 0
            \end{array} \right),
  ~~~ t^2= \frac{1}{\sqrt{2}}\left(
            \begin{array}{ccc}
                  0 & -i & 0 \\
                  i & 0 & -i \\
                  0 & i & 0
            \end{array} \right),
  ~~~ t^3=  \left(
            \begin{array}{ccc}
                  1 & 0 & 0 \\
                  0 & 0 & 0 \\
                  0 & 0 & -1
            \end{array} \right)
\end{eqnarray}
are the $SU(2)$ generators for the triplet representation, and the matrix $P$ is given by
\begin{eqnarray}
      P= \frac{1}{\sqrt{2}}
         \left(
         \begin{array}{ccc}
         -1 & i & 0 \\
         0 & 0 & \sqrt{2} \\
         1 & i & 0
         \end{array}
         \right).
\end{eqnarray}
The vacuum expectation values are defined by
\begin{eqnarray}
\langle \, \Phi \, \rangle = \frac{v_{\phi}}{\sqrt{2}}\, {\rm diag}\left( 1,~ 1 \right),~~~~~~
\langle \, \Delta \, \rangle = {\rm diag}\left( v_{\chi},~ v_{\xi},~ v_{\chi} \right).
\end{eqnarray}
The $SU(2)_L\times SU(2)_R \rightarrow SU(2)_V$ symmetry breaking implies that $v_{\chi} = v_{\xi} \equiv v_{\Delta}$ and $v_{\phi}^2 + 8v_{\Delta}^2 = v^2 \approx (246~{\rm GeV})^2$. The physical Higgs components can be organized into a fiveplet $(H_5^{\pm\pm}, H_5^{\pm}, H_5^0)$, a triplet $(H_3^{\pm}, H_3^0)$, and two singlets $h^0$ and $H^0$ of the custodial $SU(2)_V$ symmetry. These mass eigenstates are given by \cite{test-custodial}
\begin{eqnarray}
&&
H_5^{\pm \pm}=\chi^{\pm \pm},~~~~~~~~~~
H_5^{\pm}=\frac{\left( \chi^{\pm} - \xi^{\pm} \right)}{\sqrt{2}},~~~~~~~~~~
H_5^0=\sqrt{\frac{1}{3}} \, \chi^0_{{\rm R}} - \sqrt{\frac{2}{3}} \, \xi^0_{{\rm R}}, \nonumber \\
&&
H_3^{\pm}=-\sin\theta_H \, \phi^{\pm} + \cos\theta_H \, \frac{\left( \chi^{\pm} + \xi^{\pm} \right)}{\sqrt{2}},~~~~~~~~~~\,
H_3^0=-\sin\theta_H \, \phi^0_{{\rm I}} + \cos\theta_H \, \chi^0_{{\rm I}}, \\
&&
h^0=\cos\alpha \, \phi^0_{{\rm R}} - \sin\alpha \, \Big( \sqrt{\frac{2}{3}} \, \chi^0_{{\rm R}} + \sqrt{\frac{1}{3}} \, \xi^0_{{\rm R}} \Big),~~~~~~~
H^0=\sin\alpha \, \phi^0_{{\rm R}} + \cos\alpha \, \Big( \sqrt{\frac{2}{3}} \, \chi^0_{{\rm R}} + \sqrt{\frac{1}{3}} \, \xi^0_{{\rm R}} \Big), \nonumber
\end{eqnarray}
where $\cos\theta_H = v_{\phi}/v$, and the real fields $\phi^0_{{\rm R, \, I}}$, $\chi^0_{{\rm R, \, I}}$, and $\xi^0_{{\rm R}}$ are defined by
\begin{eqnarray}
\phi^0 = \frac{v_{\phi}}{\sqrt{2}} + \frac{\phi^0_{{\rm R}} + i \phi^0_{{\rm I}}}{\sqrt{2}},~~~~~~~~~
\chi^0 = v_{\chi} + \frac{\chi^0_{{\rm R}} + i \chi^0_{{\rm I}}}{\sqrt{2}},~~~~~~~~~
\xi^0 = v_{\xi} + \xi^0_{{\rm R}}.
\end{eqnarray}
The masses of the $SU(2)_V$ fiveplet and triplet Higgs bosons can be, respectively, expressed as\footnote{The $SU(2)_V$ fiveplet Higgs bosons $(H_5^{\pm \pm},H_5^{\pm},H_5^0)$ have the same mass $m_5$ and the $SU(2)_V$ triplet Higgs bosons $(H_3^{\pm},H_3^0)$ have the same mass $m_3$.}
\begin{eqnarray}
&&
m_5^2 = \Big( \frac{M_1}{4 v_{\Delta}} + \frac{3}{2} \lambda_5 \Big) v^2
       + \Big(
         \frac{- 2 M_1 + 12 M_2}{v_{\Delta}} + 8 \lambda_3 - 12 \lambda_5
         \Big) v_{\Delta}^2, \nonumber \\
&&
m_3^2 = \Big( \frac{M_1}{4 v_{\Delta}} + \frac{1}{2} \lambda_5 \Big) v^2.
\end{eqnarray}
The mass-squared matrix of the $SU(2)_V$ singlet sector in the basis of $\left\{ \phi^0_{{\rm R}},~  \sqrt{\frac{2}{3}} \, \chi^0_{{\rm R}} + \sqrt{\frac{1}{3}} \, \xi^0_{{\rm R}} \right\}$ is written as
\begin{eqnarray}
M^2
=
\left(
\begin{array}{cc}
M^2_{11} & M^2_{12} \\
M^2_{12} & M^2_{22}
\end{array}
\right),
\end{eqnarray}
where
\begin{eqnarray}
&&
M^2_{11} = 8 \lambda_1 v_{\phi}^2, \nonumber \\
&&
M^2_{22} = \frac{M_1}{4 v_{\Delta}} v^2 + \Big( \frac{- 2 M_1 - 6 M_2}{v_{\Delta}} + 8 \lambda_3 + 24 \lambda_4 \Big) v_{\Delta}^2, \nonumber \\
&&
M^2_{12} = \frac{\sqrt{3}}{2} \Big( \frac{- M_1}{v_{\Delta}} + 8 \lambda_2 - 4 \lambda_5 \Big) v_{\phi} v_{\Delta}.
\end{eqnarray}
Consequently, the masses of the two $SU(2)_V$ singlet Higgs bosons $m_{h^0, H^0}$ and the mixing angle $\alpha$ are given by
\begin{eqnarray}
{\rm diag}\left( m^2_{h^0},~ m^2_{H^0} \right) \cong M^2, ~~~~~~~~~
\tan2\alpha = \frac{2 M^2_{12}}{M^2_{22} - M^2_{11}},
\end{eqnarray}
where $m_{h^0} < m_{H^0}$ and $M_{12} \sin2\alpha > 0$. $h^0$ is identified as the SM-like Higgs boson discovered at the LHC.

\par
The electroweak gauge sector and the Higgs sector of the GM model involve two and nine independent input parameters, respectively. In this paper, the 11 input parameters for these two sectors are chosen as
\begin{eqnarray}
\label{inputs-GM}
\left\{ \, G_F,~~ m_W,~~ m_Z,~~ m_{h^0},~~ m_5,~~ v_{\Delta},~~ M_1,~~ M_2,~~ \lambda_2,~~ \lambda_3,~~ \lambda_4  \, \right\}.
\end{eqnarray}
The first four SM input parameters are fixed as \cite{pdg}
\begin{eqnarray}
&&
G_F = 1.1663787 \times 10^{-5}~ {\rm GeV}^{-2}, \nonumber \\
&&
m_W = 80.325~ {\rm GeV},~~~
m_Z = 91.1876~ {\rm GeV},~~~
m_{h^0} = 125.09~ {\rm GeV}.~~~~
\end{eqnarray}
The rest of the seven input parameters are taken as \cite{h5plane}
\begin{eqnarray}
&&
M_1 = 4 \left( 1 + \sqrt{2} G_F m_5^2 \right) v_{\Delta},~~~~~
M_2 = \frac{1}{6} M_1, \nonumber \\
&&
\lambda_2 = 0.4 \times \frac{m_5}{1000~{\rm GeV}},~~~~~
\lambda_3 = -0.1,~~~~~
\lambda_4 = 0.2,~~~~~~~
\end{eqnarray}
where $m_5$ and $v_{\Delta}$ scan over the region of
\begin{eqnarray}
\label{m5-v}
(m_5,~ v_{\Delta}) \, \in \, [ 200,~ 1000 ] \times (0,50]~~~ ({\rm GeV}).
\end{eqnarray}
This input parameter setting, called the ``H5plane'' benchmark scenario, can be implemented by using the input set 4 in {\sc gmcalc} and satisfies the theoretical constraints from perturbative unitarity, the bounded-from-below requirement on the Higgs potential and the avoidance of $SU(2)_V$-breaking vacua \cite{decouple-limit,GM-calc}.

\par
\section{NUMERICAL RESULTS AND DISCUSSION}
\begin{figure*}[htbp]
\begin{center}
\includegraphics[scale=0.35]{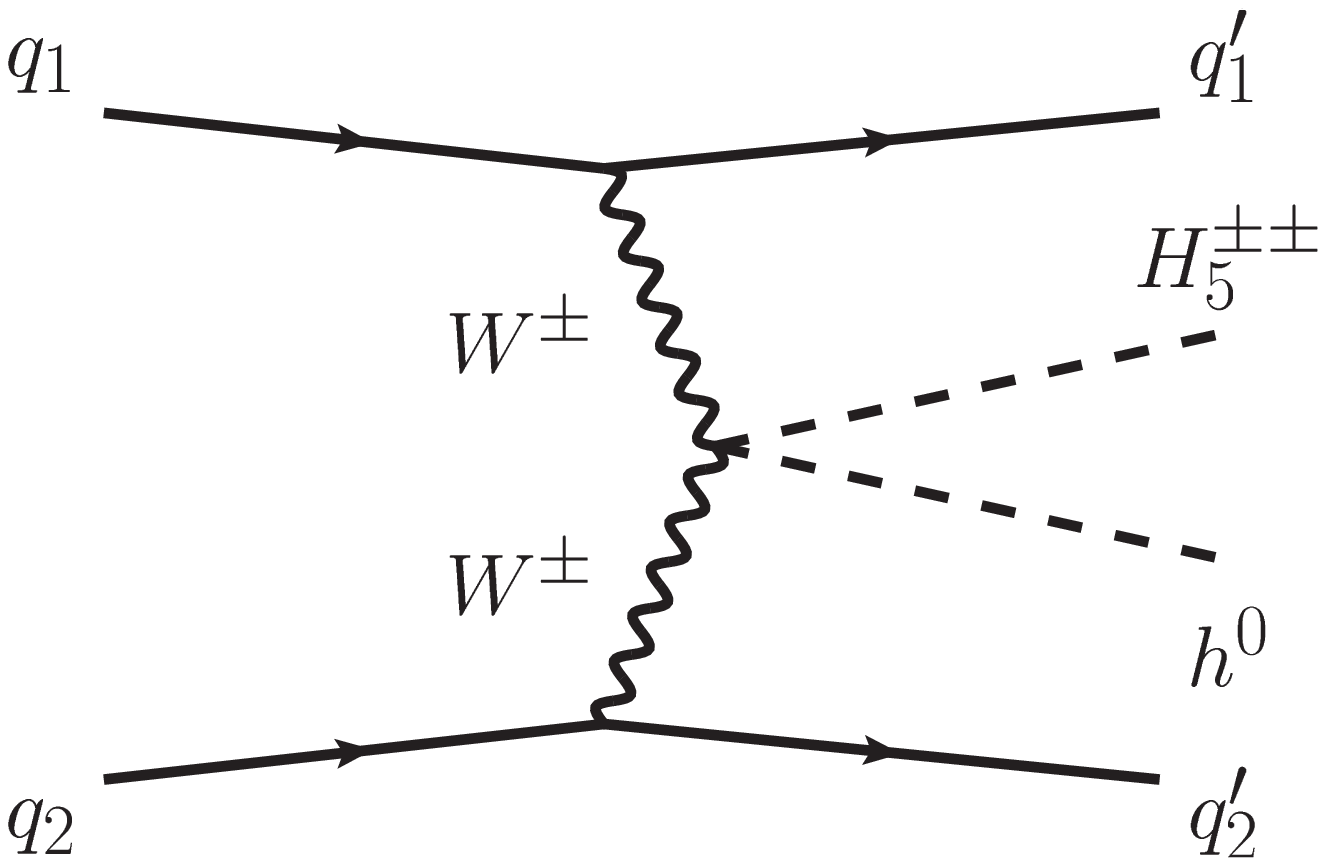}
\vspace{0.5cm}
\includegraphics[scale=0.35]{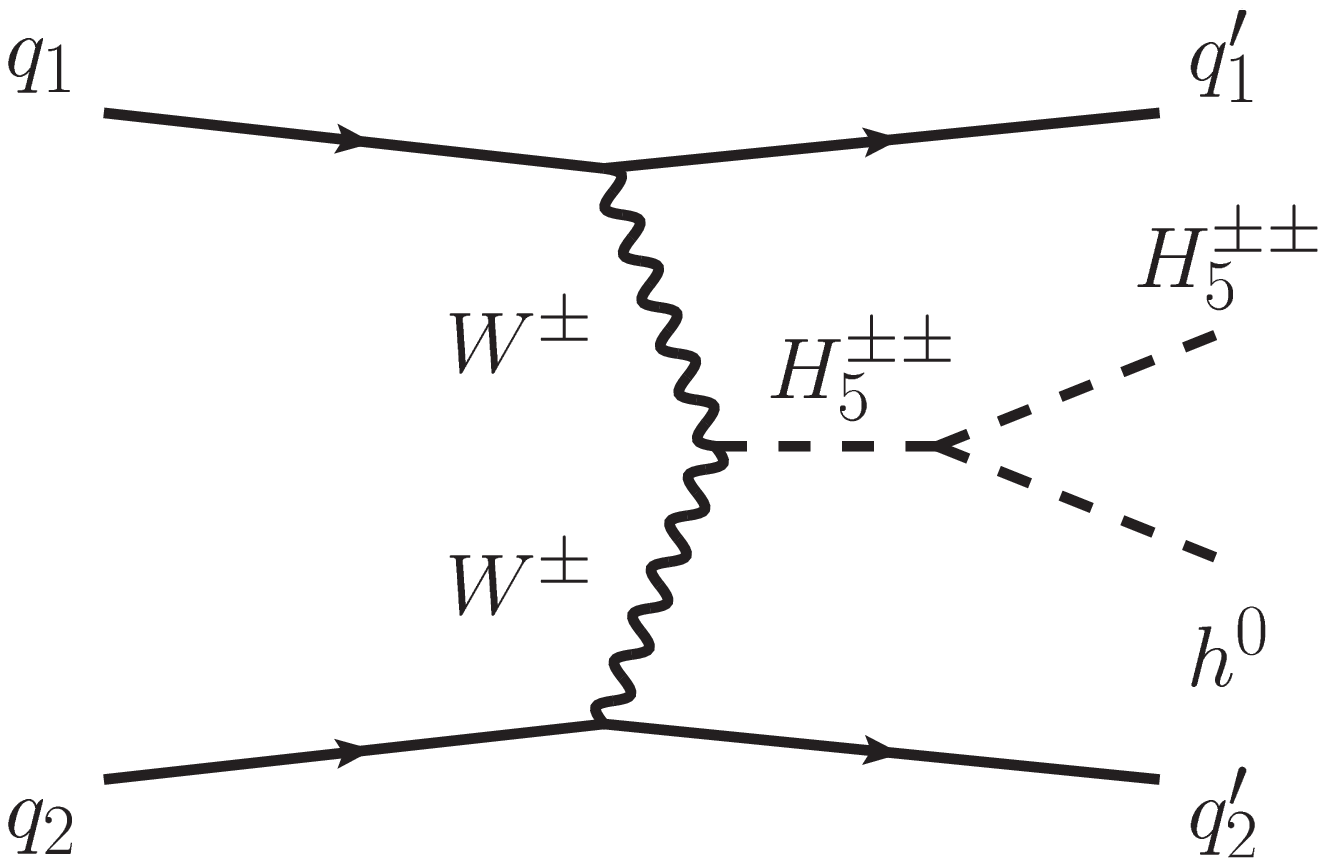}
\vspace{0.5cm}
\includegraphics[scale=0.35]{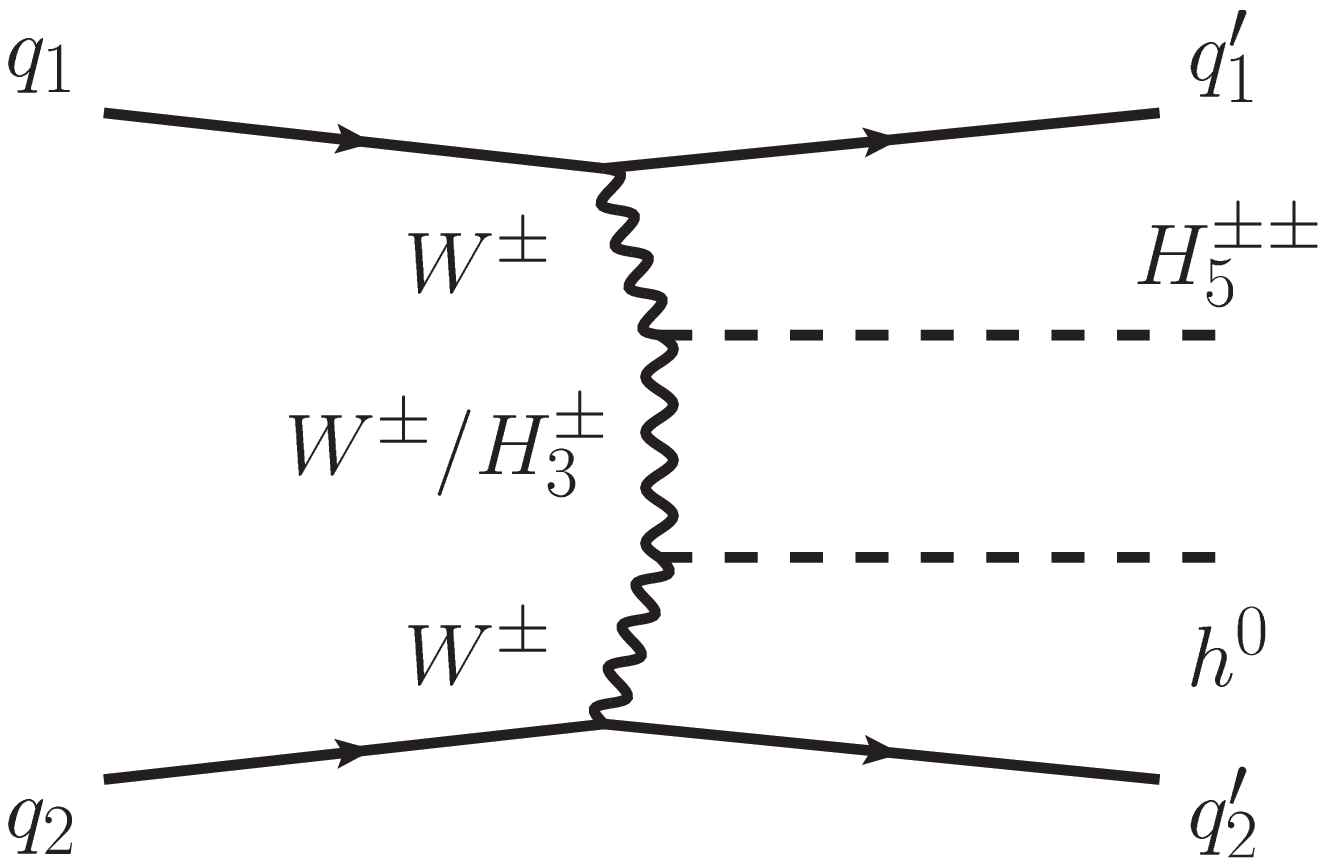}
\caption{\small LO Feynman diagrams for $q_1q_2 \rightarrow H_5^{\pm\pm} h^0 q_1^{\prime} q_2^{\prime}$ via VBF.}
\label{Feyndiag}
\end{center}
\end{figure*}

To select the VBF events characterized by two hard jets in the forward and backward rapidity regions, we apply the following VBF cuts on the final state,
\begin{eqnarray}
\label{VBFcuts}
p_{T,j_{1,2}} > 30~ {\rm GeV},~~~
|\eta_{j_{1,2}}| < 4.5,~~~
|\eta_{j_1} - \eta_{j_2}| > 4,~~~
\eta_{j_1} \cdot \eta_{j_2} < 0,~~~
M_{j_1j_2} > 600~ {\rm GeV},
\end{eqnarray}
where $j_1$ and $j_2$ represent the two hardest jets in the final state\footnote{$j_1$ and $j_2$ are called leading and next-to-leading jets, respectively, according to their transverse momentum in decreasing order, i.e., $p_{T,j_1} > p_{T,j_2}$.}, and $p_{T_{j_{1,2}}}$, $\eta_{j_{1,2}}$, and $M_{j_1 j_2}$ denote the transverse momenta, pseudorapidities and invariant mass of the two hardest jets, respectively. Compared to the VBF mechanism, the Higgs-strahlung mechanism for $H_5^{\pm\pm} h^0$-associated production, i.e., $pp \rightarrow W^{\pm} \rightarrow H_5^{\pm\pm} h^0 + 2\, {\rm jets}$, is heavily suppressed by the VBF cuts and thus can be neglected. The leading-order parton-level Feynman diagrams for $H_5^{\pm\pm} h^0$-associated production via VBF mechanism in the GM model at a hadron collider are presented in Fig.\ref{Feyndiag}. In this section we study in detail the associated production of $H_5^{\pm\pm} h^0$ via the VBF mechanism at the $14~ {\rm TeV}$ LHC and $70~ {\rm TeV}$ Super Proton-Proton Collider (SPPC) \cite{sppc}. We use the {\sc feynrules} \cite{FeynRules} and {\sc nloct} \cite{NLOCT} packages to generate a model file in Universal FeynRules Output format \cite{UFO} at the QCD NLO, and then employ {\sc MadGraph5\_aMC@NLO} to calculate the NLO QCD corrections in the GM model. We adopt the NNPDF2.3QED parton distribution functions in the initial-state parton convolution, set the factorization and renormalization scales to the paronic colliding energy in the center-of-mass frame ($\mu_{\rm R} = \mu_{\rm F} = \sqrt{\hat{s}}$), assume $V_{{\rm CKM}} = 1_{3 \times 3}$ since the VBF production rate is independent of quark mixing, and use the anti-$k_T$ algorithm \cite{anti-kt} with radius $\Delta R = 0.4$ to cluster hadrons into jets with the help of {\sc fastjet} \cite{fastjet1,fastjet2} in NLO QCD real emission corrections.

\subsection{$H^{\pm\pm} h^0$ VBF production}
\begin{figure*}[htbp]
\begin{center}
\includegraphics[scale=0.48]{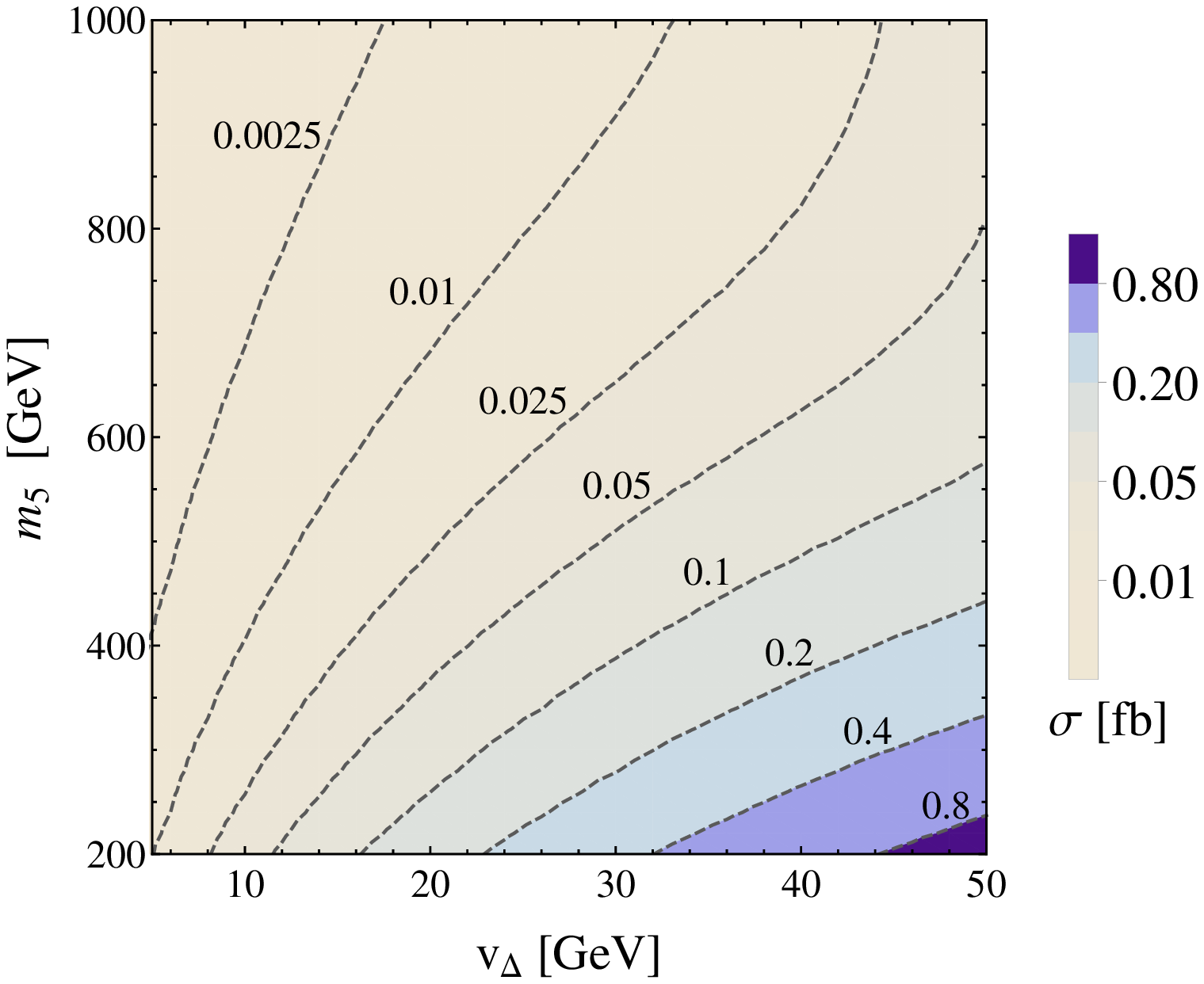}
\includegraphics[scale=0.48]{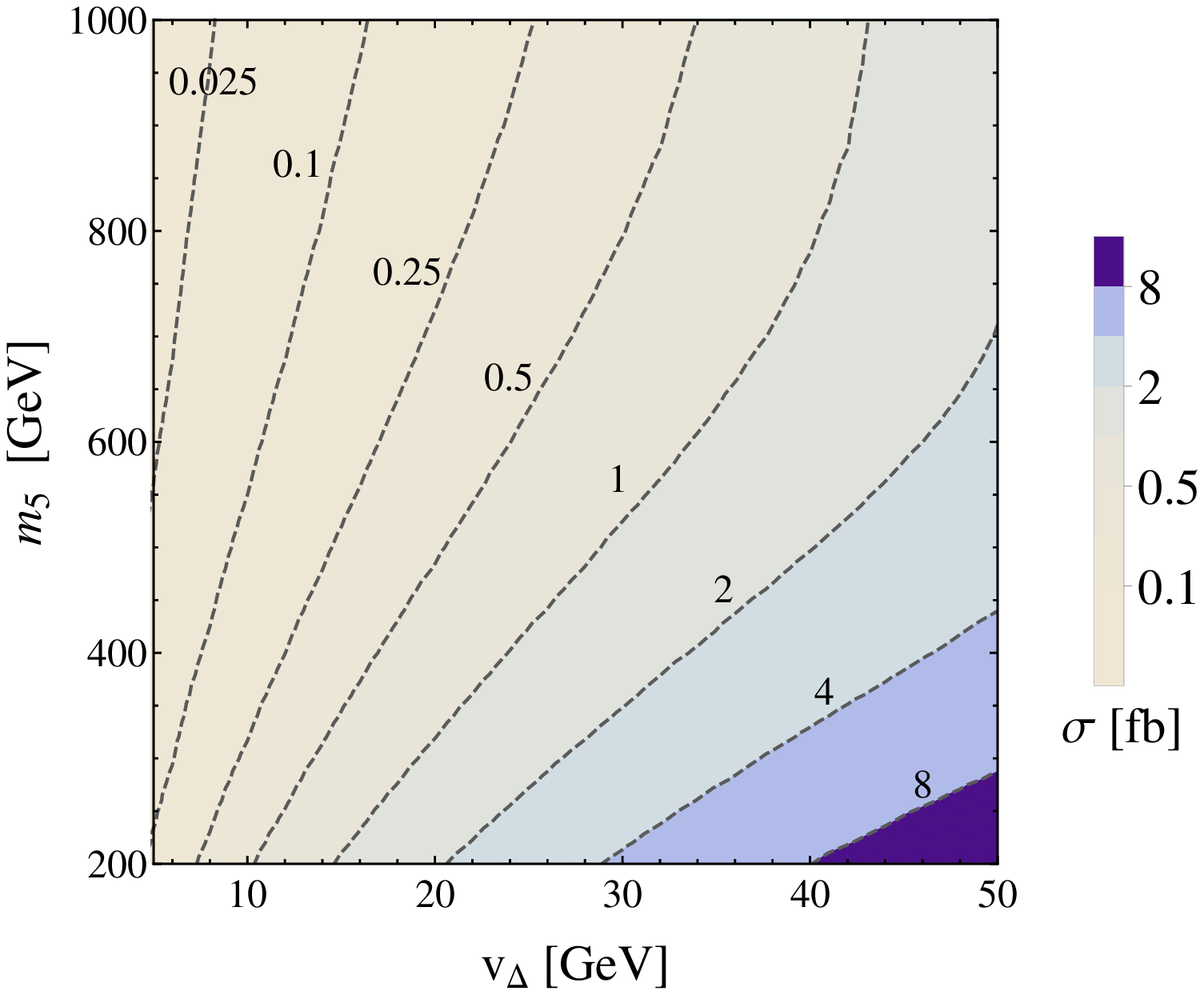}
\caption{\small
Contours of $\sigma(pp \rightarrow W^{+}W^{+} \rightarrow H_5^{++} h^0 + 2\, {\rm jets})$ in the H5plane benchmark scenario of the GM model at the $14~ {\rm TeV}$ LHC (left) and $70~ {\rm TeV}$ SPPC (right).
}
\label{xsection}
\end{center}
\end{figure*}

In Fig.\ref{xsection}, we display the leading-order (LO) cross section for $pp \rightarrow W^{+}W^{+} \rightarrow H_5^{++} h^0 + 2\, {\rm jets}$ as a function of $m_5$ and $v_{\Delta}$ in the H5plane benchmark scenario of the GM model at the $14~ {\rm TeV}$ LHC (left) and $70~ {\rm TeV}$ SPPC (right), respectively. As shown in this figure, the line shapes of the contours for different colliding energies are very similar. The production cross section increases as the increment of $v_{\Delta}$ due to the enhancement of self- and gauge couplings of fiveplet Higgs bosons, while it decreases as the increment of $m_5$ due to the suppression of final-state phase space. As we expect, the dependence of the production rate on $m_5$ at the $70~ {\rm TeV}$ SPPC is much weaker than that at the $14~ {\rm TeV}$ LHC, especially in the small $v_{\Delta}$ region, because the phase-space suppression is not noticeable at very high-energy colliders. In the following, we discuss only the four benchmark points listed in Table \ref{benchmarkABCD} in the H5plane benchmark scenario of the GM model. These benchmark points satisfy not only the theoretical constraints mentioned before, but also the direct experimental constraint from the search for $H^{\pm\pm}_5$ via VBF at the LHC \cite{direct-constraints}, as well as the indirect experimental constrains from $B$ physics (such as $b \rightarrow s \gamma$, $B^0-\bar{B}^0$ mixing, and $R_b$) and electroweak oblique parameters \cite{indirect-constraints}.
\begin{table}[htbp]
\begin{center}
\renewcommand\arraystretch{1.5}
\begin{tabular}{|c||c|c|c|c|}
\hline
Benchmark point   &       A      &       B      &       C      &       D      \\
\hline
$(m_5, v_\Delta)~ ({\rm GeV})$  &  $(200, 20)$  &  $(200, 17)$  &  $(300, 20)$  &  $(300, 17)$  \\
\hline
\end{tabular}
\caption{\small Four benchmark points in the H5plane benchmark scenario of the GM model.}
\label{benchmarkABCD}
\end{center}
\end{table}

\par
The Higgs boson production via VBF at hadron colliders has been widely investigated up to the QCD NNLO in both the SM \cite{vbf-nnlo,sf-lls} and the GM model \cite{gm-sf-nnlo} by using the structure function technique. However, within the framework of structure function, we cannot study the kinematic distributions of final jets beyond the LO because the final jets can only be well defined at the LO. Thus a full perturbative calculation (simulation) is necessary when considering the NLO QCD corrections to jet distributions.

\par
In Tables \ref{tableH5++} and \ref{tableH5--}, we present the LO and NLO QCD corrected integrated cross sections for the VBF processes $pp \rightarrow W^{+}W^{+} \rightarrow H_5^{++} h^0 + 2\, {\rm jets}$ and $pp \rightarrow W^{-}W^{-} \rightarrow H_5^{--} h^0 + 2\, {\rm jets}$ at the $14$ and $70~ {\rm TeV}$ $pp$ colliders, respectively, where the scale uncertainties are calculated by varying the renormalization and factorization scales simultaneously in the range of $[ \sqrt{\hat{s}}/2,~ 2 \sqrt{\hat{s}} ]$. The two tables show that the cross section at benchmark point A is about $1.3 \sim 1.9$ times larger than that at benchmark point D. At the $14~ {\rm TeV}$ LHC, the scale uncertainties are reduced significantly by the NLO QCD correction. The NLO QCD relative correction is almost independent of $m_5$ and $v_{\Delta}$, and the QCD $K$ factors for $H_5^{++} h^0$ and $H_5^{--} h^0$ VBF productions are about $1.21$ and $1.25$, respectively. At the $70~ {\rm TeV}$ SPPC, the LO scale uncertainties are less than $1\%$, while the QCD NLO scale uncertainties are about $3\%$. The production cross section is of ${\cal O}(1)~ fb$, and the NLO QCD relative correction is more sensitive to $m_5$ compared to that at the $14~ {\rm TeV}$ LHC. As the increment of $m_5$ from $200$ to $300~ {\rm GeV}$, the QCD $K$ factors for $H_5^{++} h^0$ and $H_5^{--} h^0$ VBF productions increase from $1.03$ to $1.05$ and from $1.10$ to $1.13$, respectively. At both the $14~ {\rm TeV}$ LHC and $70~ {\rm TeV}$ SPPC, the VBF production of $H_5^{--} h^0$ receives slightly larger QCD correction compared to the VBF production of $H_5^{++} h^0$, which is consistent with the production of the single doubly charged Higgs boson via VBF in Ref.\cite{automatic-nlo}.
\begin{table}[htbp]
\begin{center}
\renewcommand\arraystretch{1.5}
\begin{tabular}{|c||c|r|r|c|}
\hline
Benchmark point &   $\sqrt{S}$ (TeV)  &   $\sigma_{\rm LO}$ (fb)~~~   &   $\sigma_{\rm NLO}$ (fb)~~   &   $K$   \\
\hline
\multirow{2}{*}{A}
& 14 &   $ 0.1522^{\, +6.6\%}_{\, -5.9\%}   $   &   $ 0.1826^{\, +1.8\%}_{\, -2.1\%}  $   &   $ 1.20 $   \\
& 70 &   $ 1.876^{\, +0.3\%}_{\, -0.4\%}    $   &   $ 1.931^{\, +3.2\%}_{\, -2.8\%}   $   &   $ 1.03 $   \\ \cline{1-5}
\multirow{2}{*}{B}
& 14 &   $ 0.1095^{\, +6.3\%}_{\, -6.0\%}   $   &   $ 0.1320^{\, +1.7\%}_{\, -2.1\%}  $   &   $ 1.21 $   \\
& 70 &   $ 1.348^{\, +0.4\%}_{\, -0.2\%}    $   &   $ 1.386^{\, +3.2\%}_{\, -2.8\%}   $   &   $ 1.03 $   \\ \cline{1-5}
\multirow{2}{*}{C}
& 14 &   $ 0.07642^{\, +7.1\%}_{\, -6.4\%}  $   &   $ 0.09265^{\, +2.2\%}_{\, -2.4\%} $   &   $ 1.21 $   \\
& 70 &   $ 1.098^{\, +0.2\%}_{\, -0.1\%}    $   &   $ 1.155^{\, +3.1\%}_{\, -2.5\%}   $   &   $ 1.05 $   \\ \cline{1-5}
\multirow{2}{*}{D}
& 14 &   $ 0.05507^{\, +7.1\%}_{\, -6.4\%}  $   &   $ 0.06675^{\, +2.2\%}_{\, -2.5\%} $   &   $ 1.21 $   \\
& 70 &   $ 0.7894^{\, +0.3\%}_{\, -0.1\%}   $   &   $ 0.8276^{\, +3.1\%}_{\, -2.6\%}  $   &   $ 1.05 $   \\ \cline{1-5}
\hline
\end{tabular}
\caption{\small LO and NLO QCD corrected integrated cross sections with the corresponding upper and lower scale uncertainties for $pp \rightarrow W^{+}W^{+} \rightarrow H_5^{++} h^0 + 2\, {\rm jets}$ at the $14$ and $70~ {\rm TeV}$ $pp$ colliders.}
\label{tableH5++}
\end{center}
\end{table}
\begin{table}[htbp]
\begin{center}
\renewcommand\arraystretch{1.5}
\begin{tabular}{|c||c|r|r|c|}
\hline
Benchmark point &   $\sqrt{S}$ (TeV)  &   $\sigma_{\rm LO}$ (fb)~~~   &   $\sigma_{\rm NLO}$ (fb)~~   &   $K$   \\
\hline
\multirow{2}{*}{A}
&    14 &   $ 0.04708^{\,+6.4\%}_{\,-5.8\%}  $   &   $ 0.05890^{\,+1.7\%}_{\,-2.2\%}  $   &   $ 1.25 $   \\
&    70 &   $ 1.094^{\,+0.4\%}_{\,-0.7\%}    $   &   $ 1.209^{\,+3.3\%}_{\,-3.0\%}    $   &   $ 1.10 $   \\ \cline{1-5}
\multirow{2}{*}{B}
&    14 &   $ 0.03386^{\,+6.4\%}_{\,-5.8\%}  $   &   $ 0.04243^{\,+1.4\%}_{\,-1.9\%}  $   &   $ 1.25 $   \\
&    70 &   $ 0.7864^{\,+0.5\%}_{\,-0.7\%}   $   &   $ 0.8685^{\,+3.2\%}_{\,-2.9\%}   $   &   $ 1.10 $   \\ \cline{1-5}
\multirow{2}{*}{C}
&    14 &   $ 0.02221^{\,+7.1\%}_{\,-6.4\%}  $   &   $ 0.02798^{\,+2.1\%}_{\,-2.5\%}  $   &   $ 1.26 $   \\
&    70 &   $ 0.6137^{\,+0.1\%}_{\,-0.1\%} $   &   $ 0.6908^{\,+3.0\%}_{\,-2.6\%}   $   &   $ 1.13 $   \\ \cline{1-5}
\multirow{2}{*}{D}
&    14 &   $ 0.01601^{\,+7.1\%}_{\,-6.4\%}  $   &   $ 0.02004^{\,+3.0\%}_{\,-3.2\%}  $   &   $ 1.25 $   \\
&    70 &   $ 0.4417^{\,+0.1\%}_{\,-0.1\%} $   &   $ 0.4950^{\,+3.0\%}_{\,-2.6\%}   $   &   $ 1.12 $   \\ \cline{1-5}
\hline
\end{tabular}
\caption{\small Same as Table \ref{tableH5++} but for $pp \rightarrow W^{-}W^{-} \rightarrow H_5^{--} h^0 + 2\, {\rm jets}$.}
\label{tableH5--}
\end{center}
\end{table}

\par
Now we turn to the differential distributions with respect to some kinematic variables of final particles. Since the differential cross sections for $H_5^{--} h^0$ VBF production are similar to those for $H_5^{++} h^0$ VBF production, we only consider $pp \rightarrow W^{+}W^{+} \rightarrow H_5^{++} h^0 + 2\, {\rm jets}$ at the $14~ {\rm TeV}$ LHC and $70~ {\rm TeV}$ SPPC at benchmark point A in the following discussion.

\par
\begin{figure}[htbp]
\begin{center}
\includegraphics[scale=0.35]{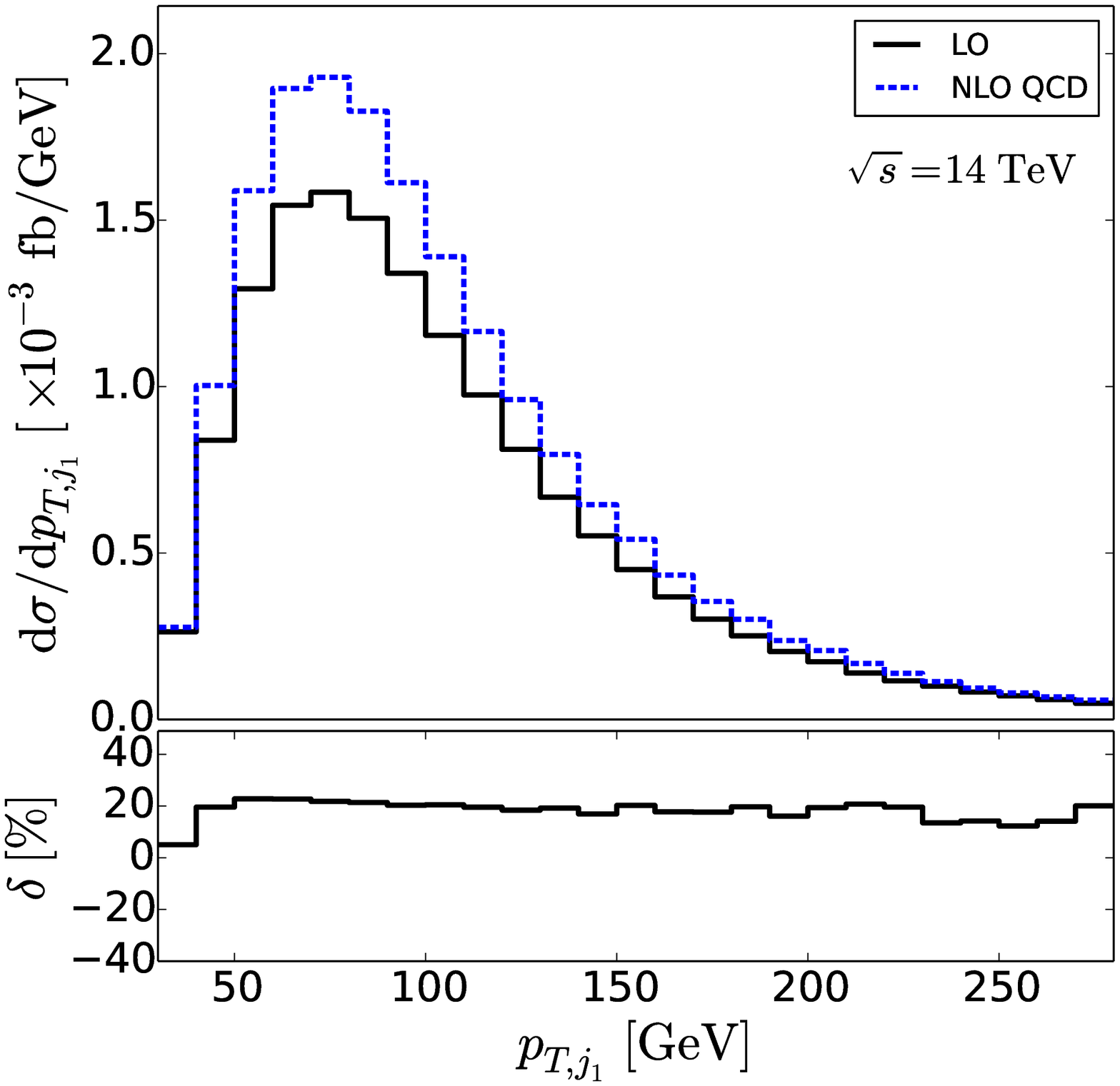}
\includegraphics[scale=0.35]{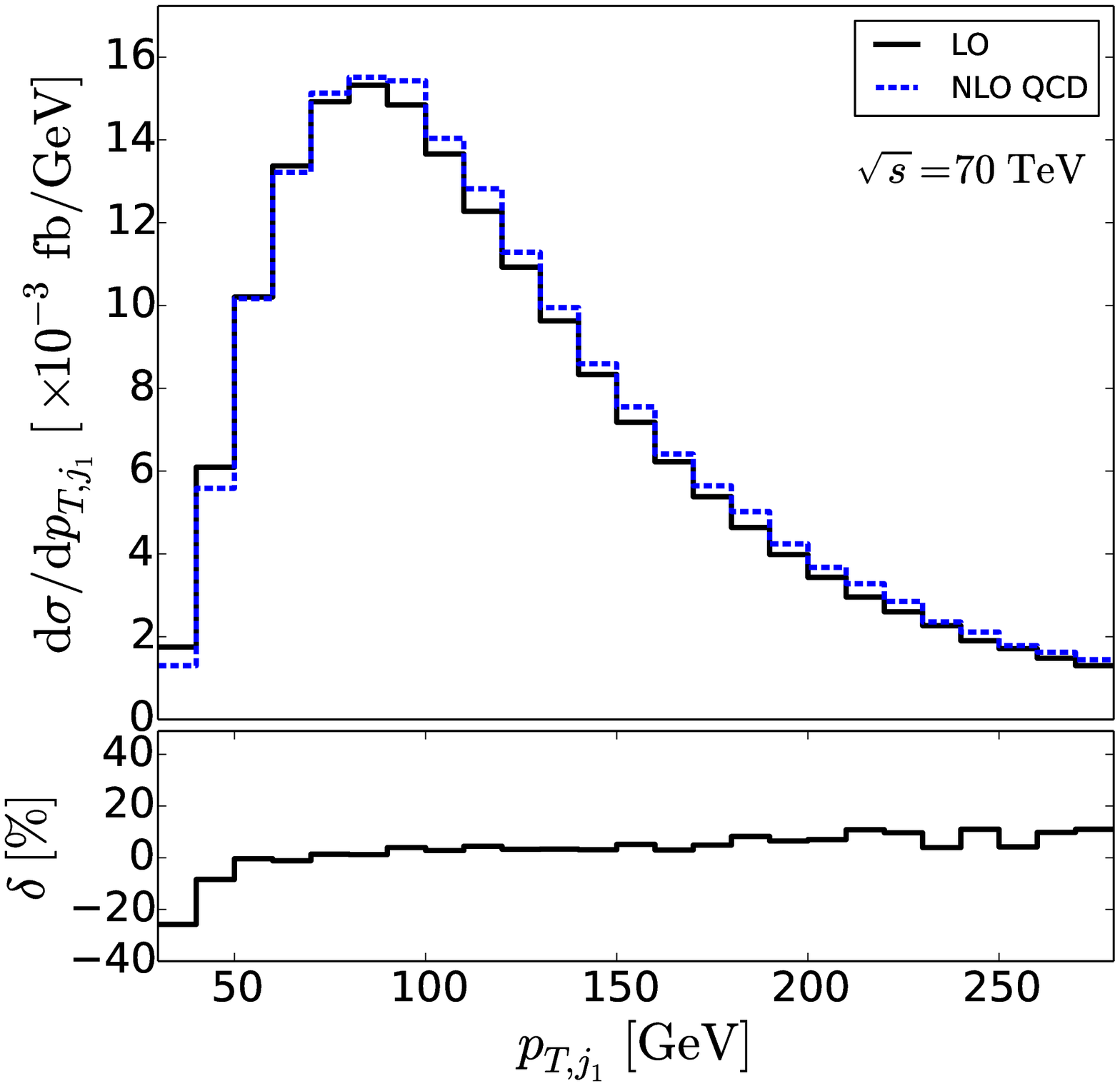}
\caption{Transverse momentum distributions of the leading jet for $pp \rightarrow W^{+}W^{+} \rightarrow H_5^{++} h^0 + 2\, {\rm jets}$ at hadron colliders.}
\label{pt-jet}
\end{center}
\end{figure}
\begin{figure}[htbp]
\begin{center}
\includegraphics[scale=0.35]{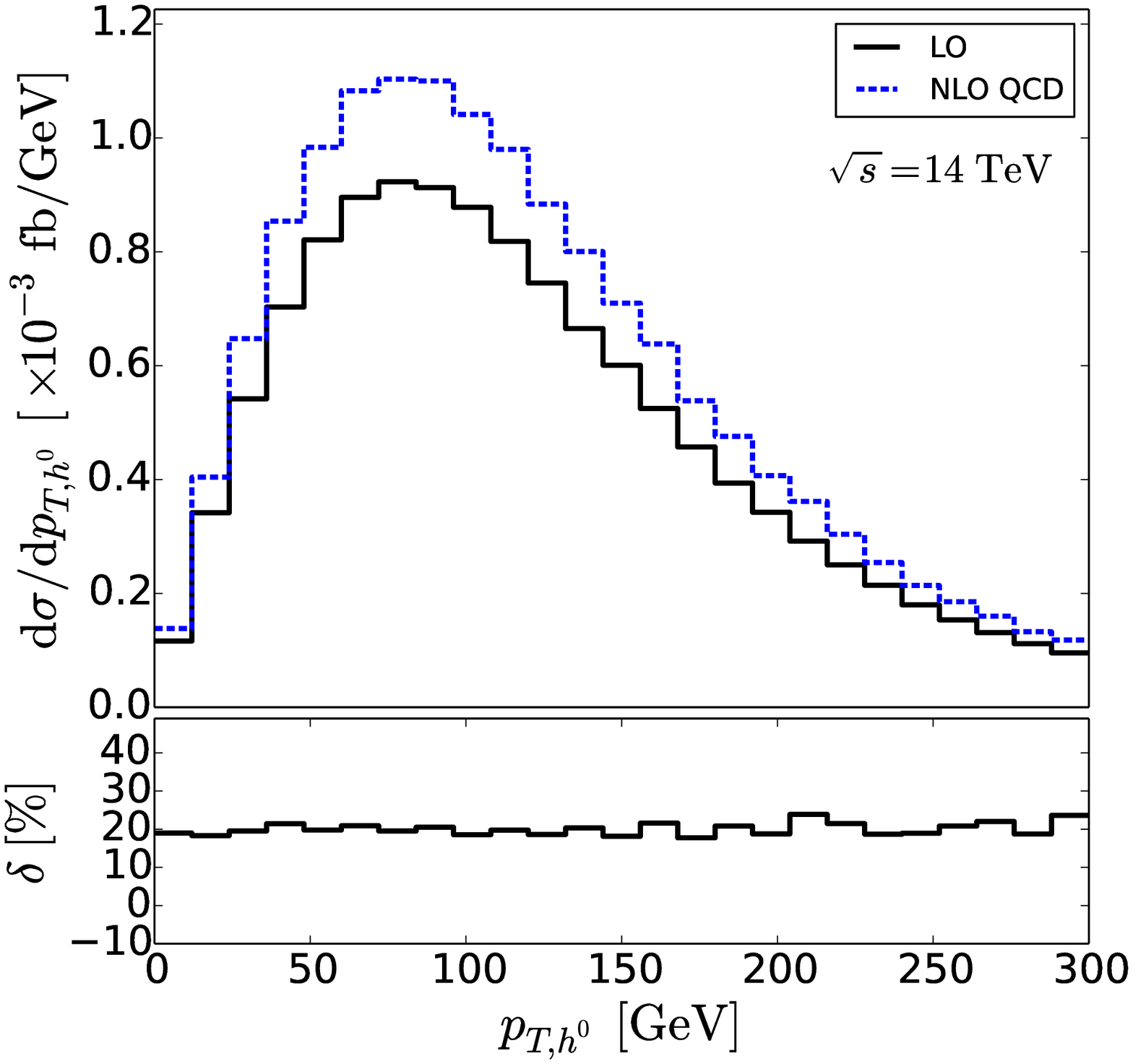}
\includegraphics[scale=0.35]{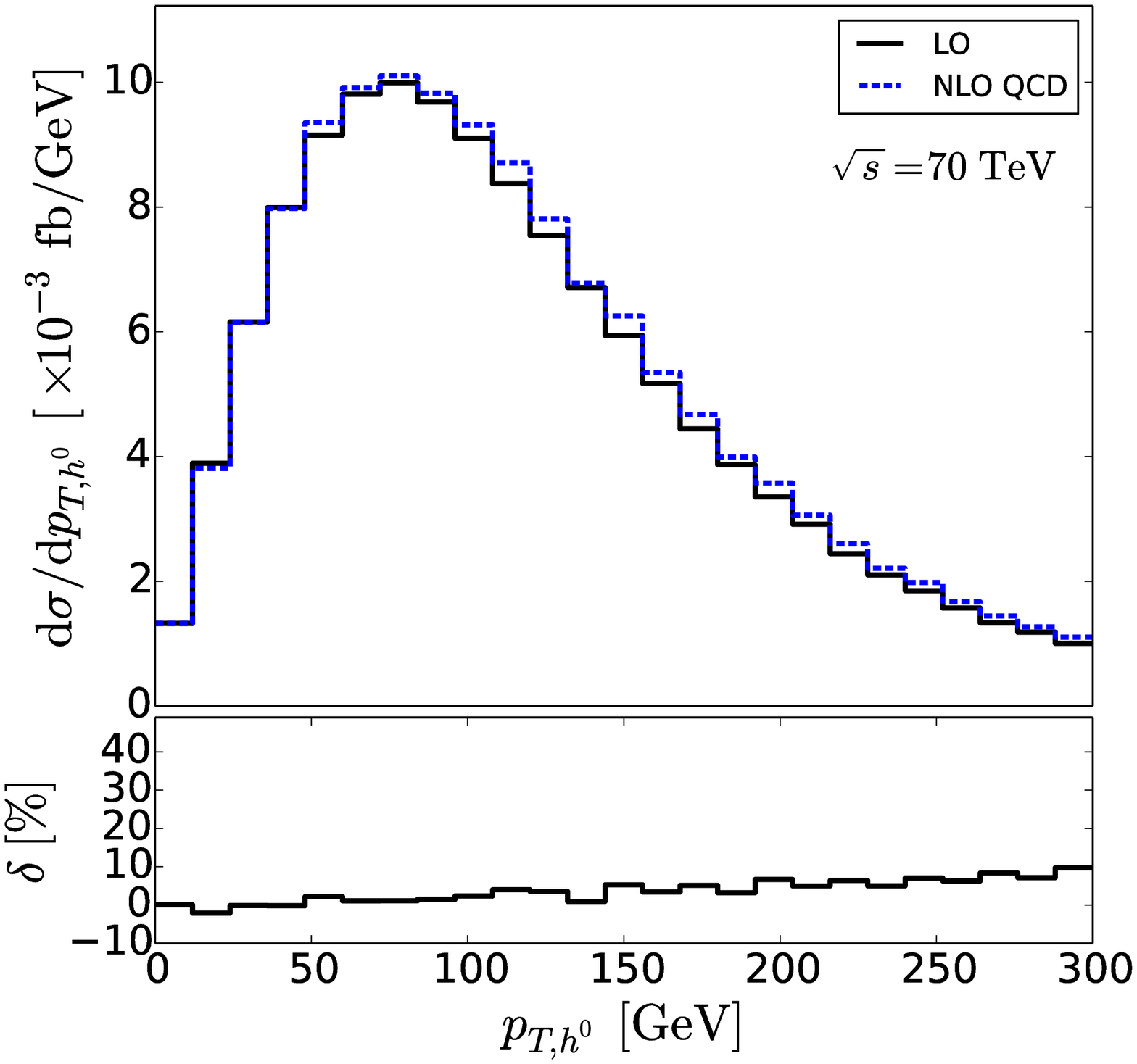}
\caption{Transverse momentum distributions of $h^0$ for $pp \rightarrow W^{+}W^{+} \rightarrow H_5^{++} h^0 + 2\, {\rm jets}$ at hadron colliders.}
\label{pt-h0}
\end{center}
\end{figure}
\begin{figure}[htbp]
\begin{center}
\includegraphics[scale=0.35]{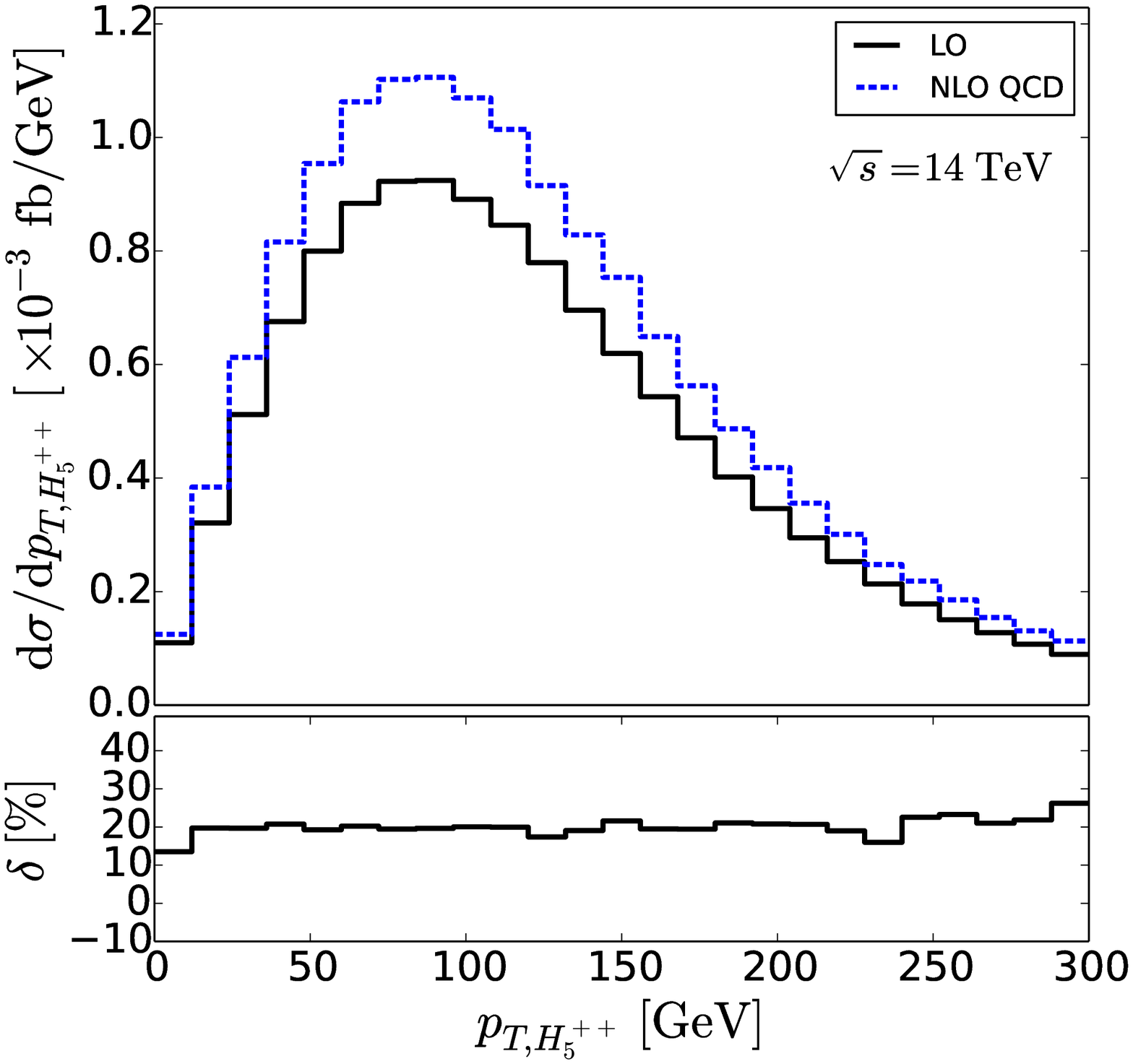}
\includegraphics[scale=0.35]{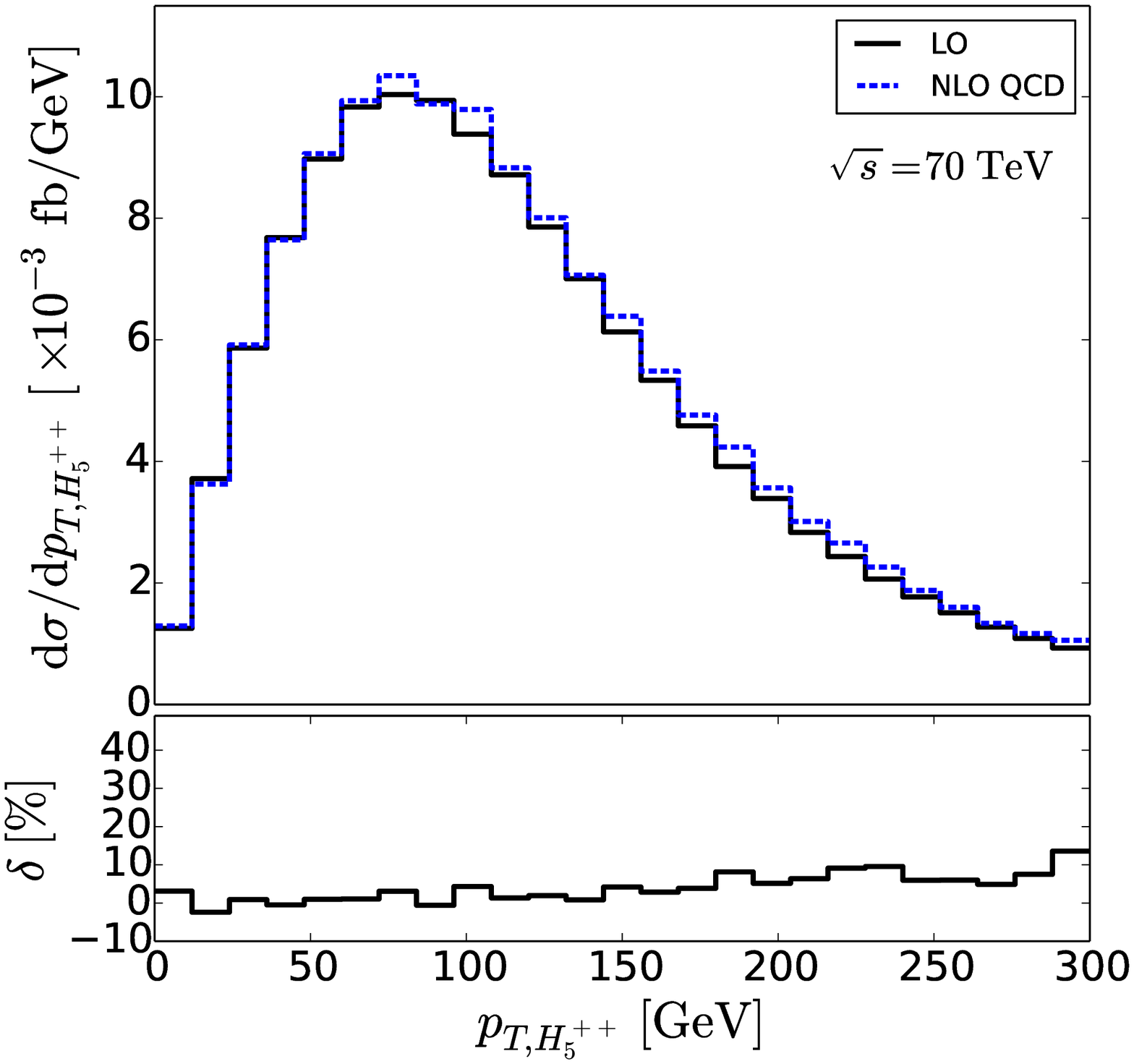}
\caption{Transverse momentum distributions of $H_5^{++}$ for $pp \rightarrow W^{+}W^{+} \rightarrow H_5^{++} h^0 + 2\, {\rm jets}$ at hadron colliders.}
\label{pt-H5}
\end{center}
\end{figure}

We present the LO and NLO QCD corrected transverse momentum distributions of the leading jet, $h^0$ and $H_5^{++}$ in Figs.\ref{pt-jet}-\ref{pt-H5}, separately. The corresponding QCD relative corrections are provided in the lower panels of these figures. We can see that all these transverse momentum distributions increase sharply in the low $p_T$ region ($p_T < 70~ {\rm GeV}$) and decrease smoothly when $p_T > 80~ {\rm GeV}$ with the increment of $p_T$. At the $14~ {\rm TeV}$ LHC, the NLO QCD correction enhances the LO transverse momentum distributions noticeably and the QCD relative corrections are steady at about $20\%$ in most of the plotted $p_T$ region. At the $70~ {\rm TeV}$ SPPC, the NLO QCD correction modifies the LO transverse momentum distributions slightly and the QCD relative corrections are positive and only about ${\cal O}(1\%)$ when $p_T > 50~ {\rm GeV}$.

\par
\begin{figure}[htbp]
\begin{center}
\includegraphics[scale=0.35]{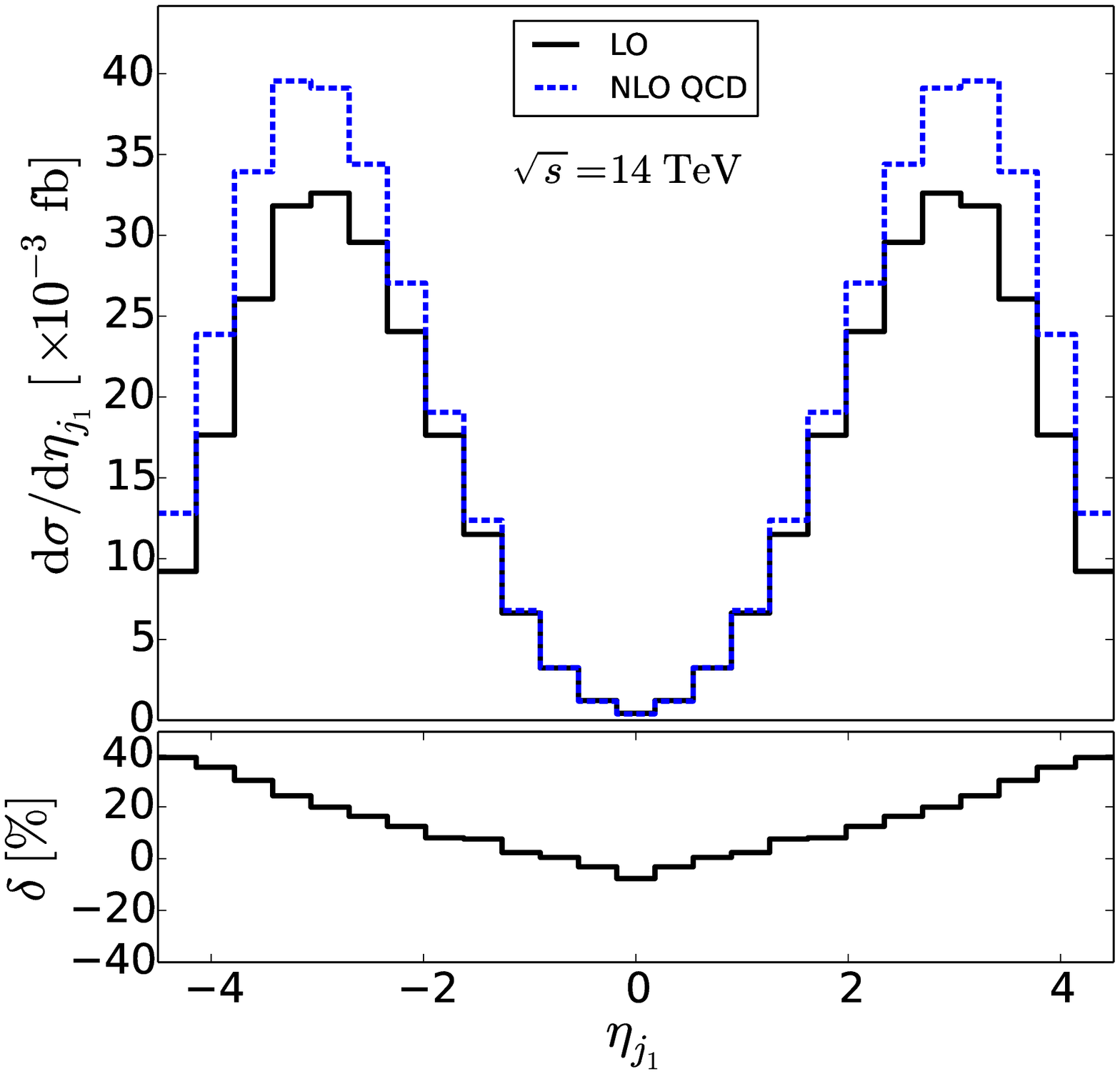}
\includegraphics[scale=0.35]{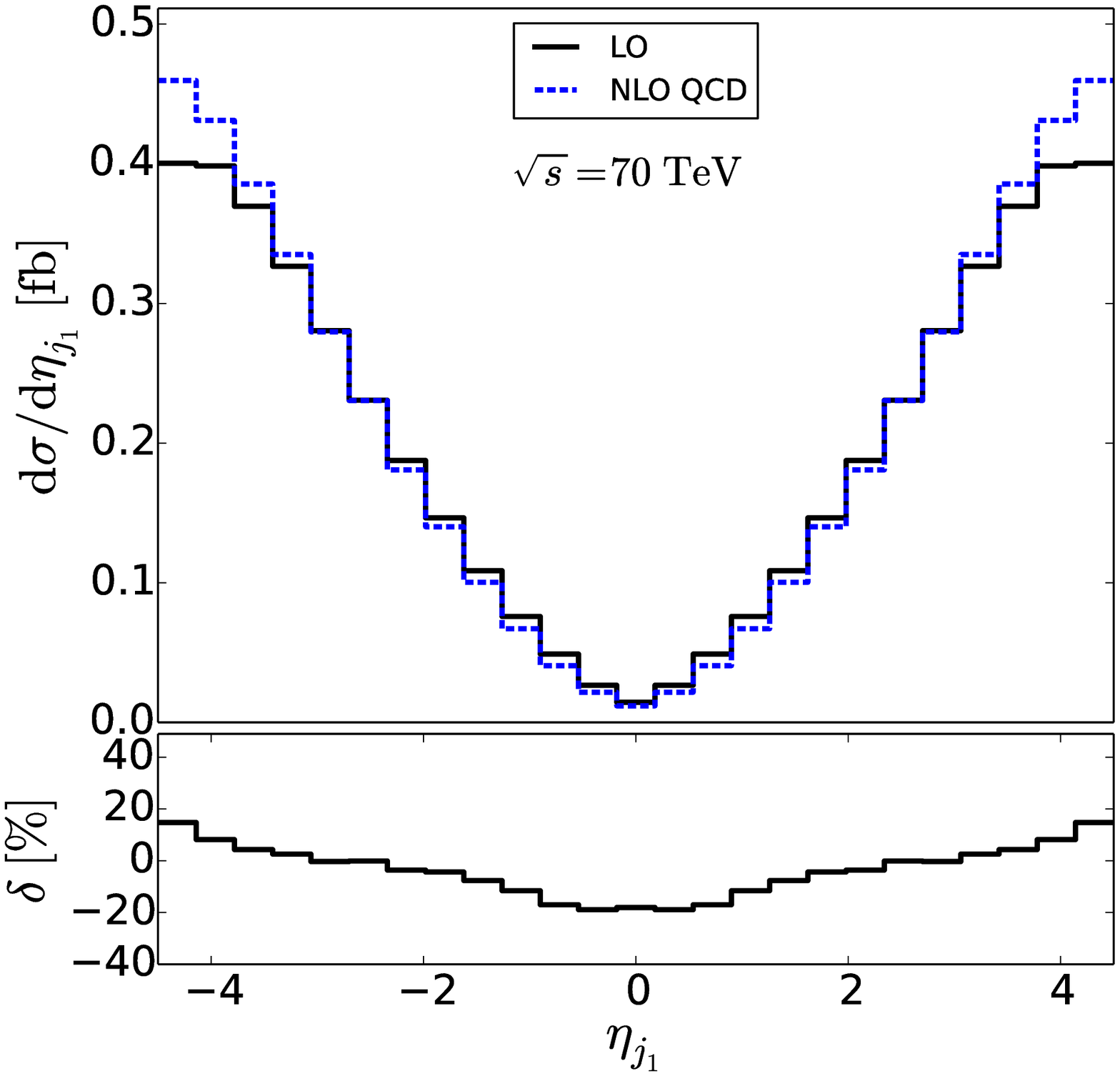}
\caption{Rapidity distributions of the leading jet for $pp \rightarrow W^{+}W^{+} \rightarrow H_5^{++} h^0 + 2\, {\rm jets}$ at hadron colliders.}
\label{rapidity-jet}
\end{center}
\end{figure}
\begin{figure}[htbp]
\begin{center}
\includegraphics[scale=0.35]{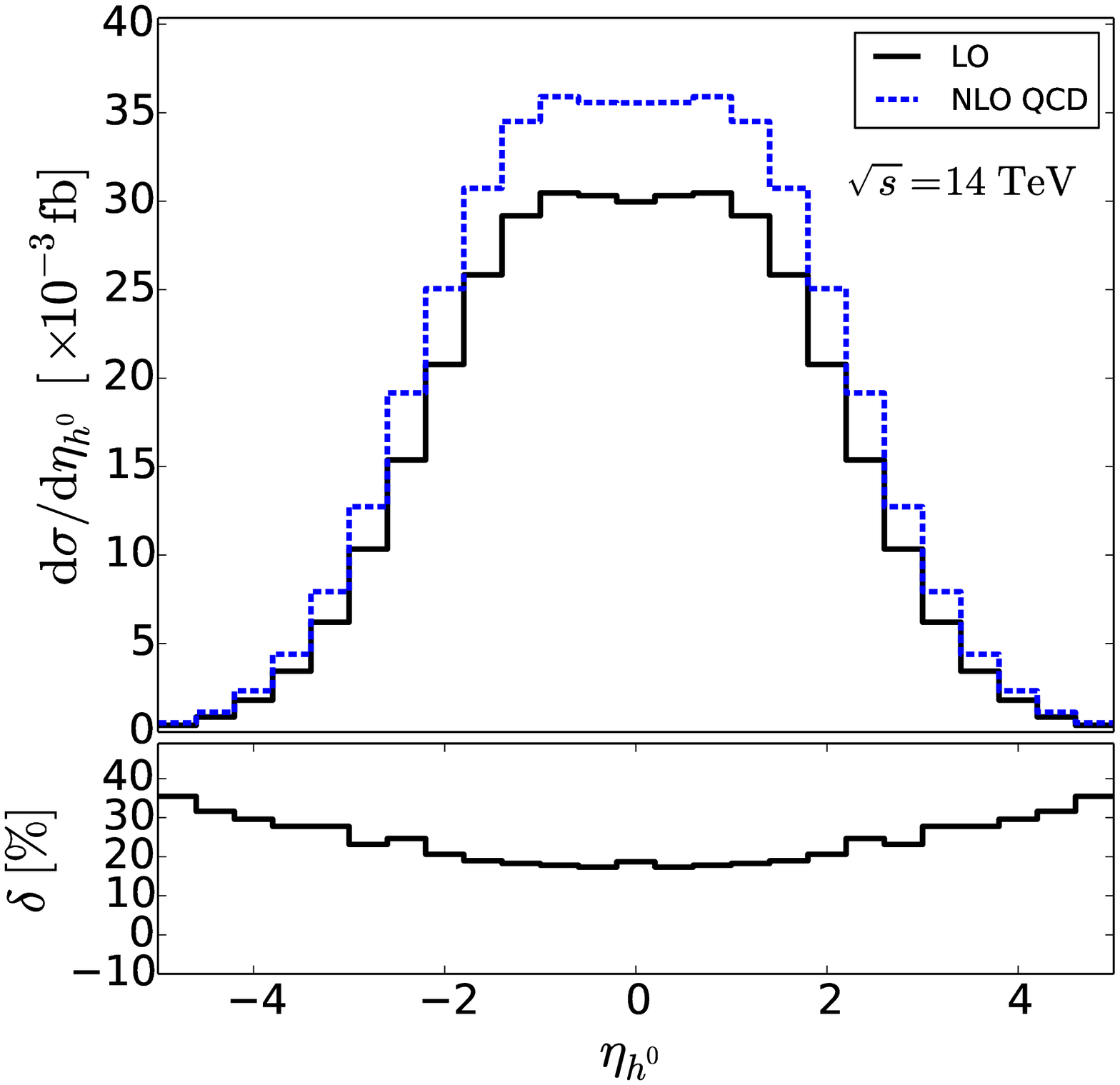}
\includegraphics[scale=0.35]{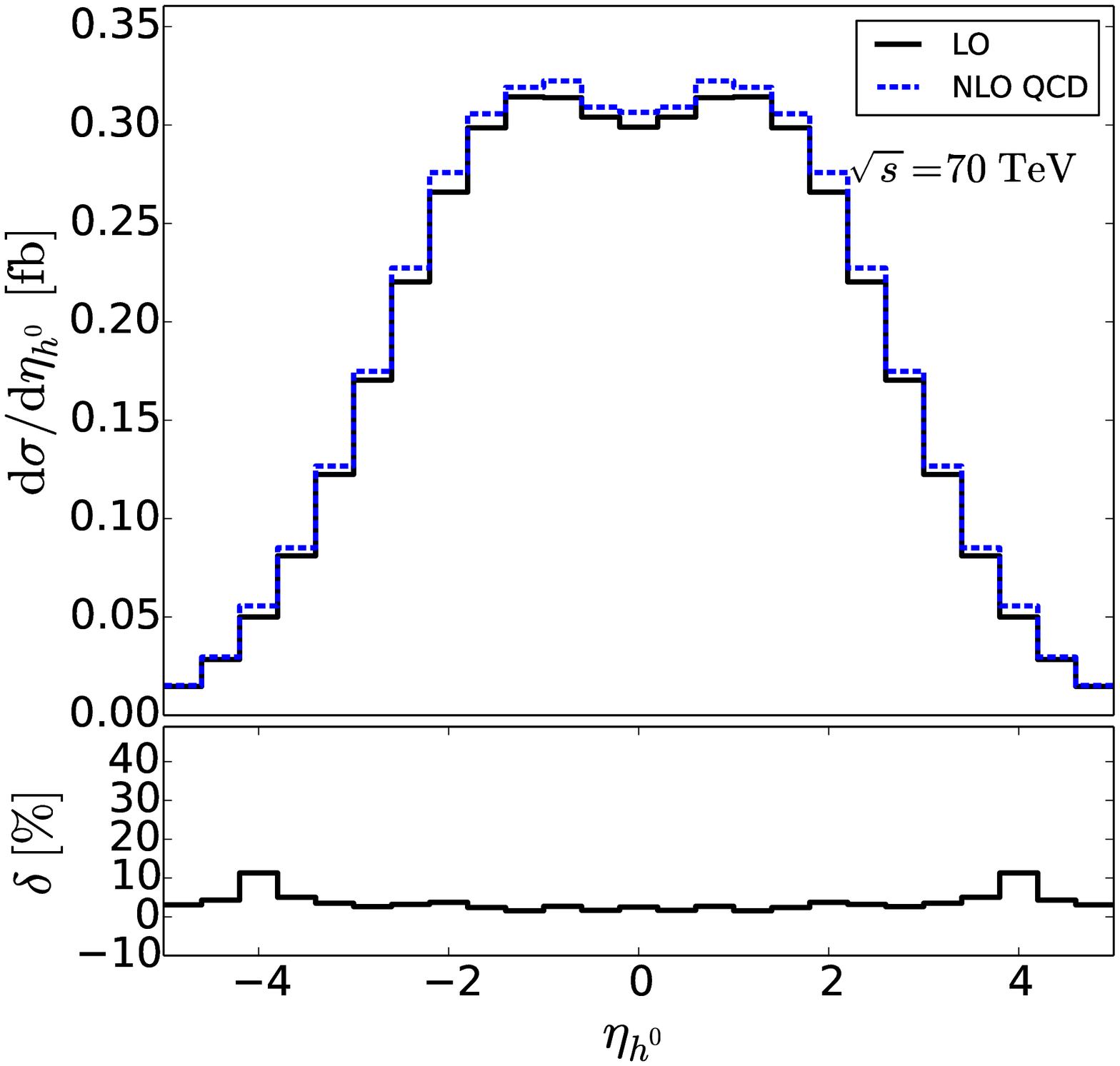}
\caption{Rapidity distributions of $h^0$ for $pp \rightarrow W^{+}W^{+} \rightarrow H_5^{++} h^0 + 2\, {\rm jets}$ at hadron colliders.}
\label{rapidity-h0}
\end{center}
\end{figure}
\begin{figure}[htbp]
\begin{center}
\includegraphics[scale=0.35]{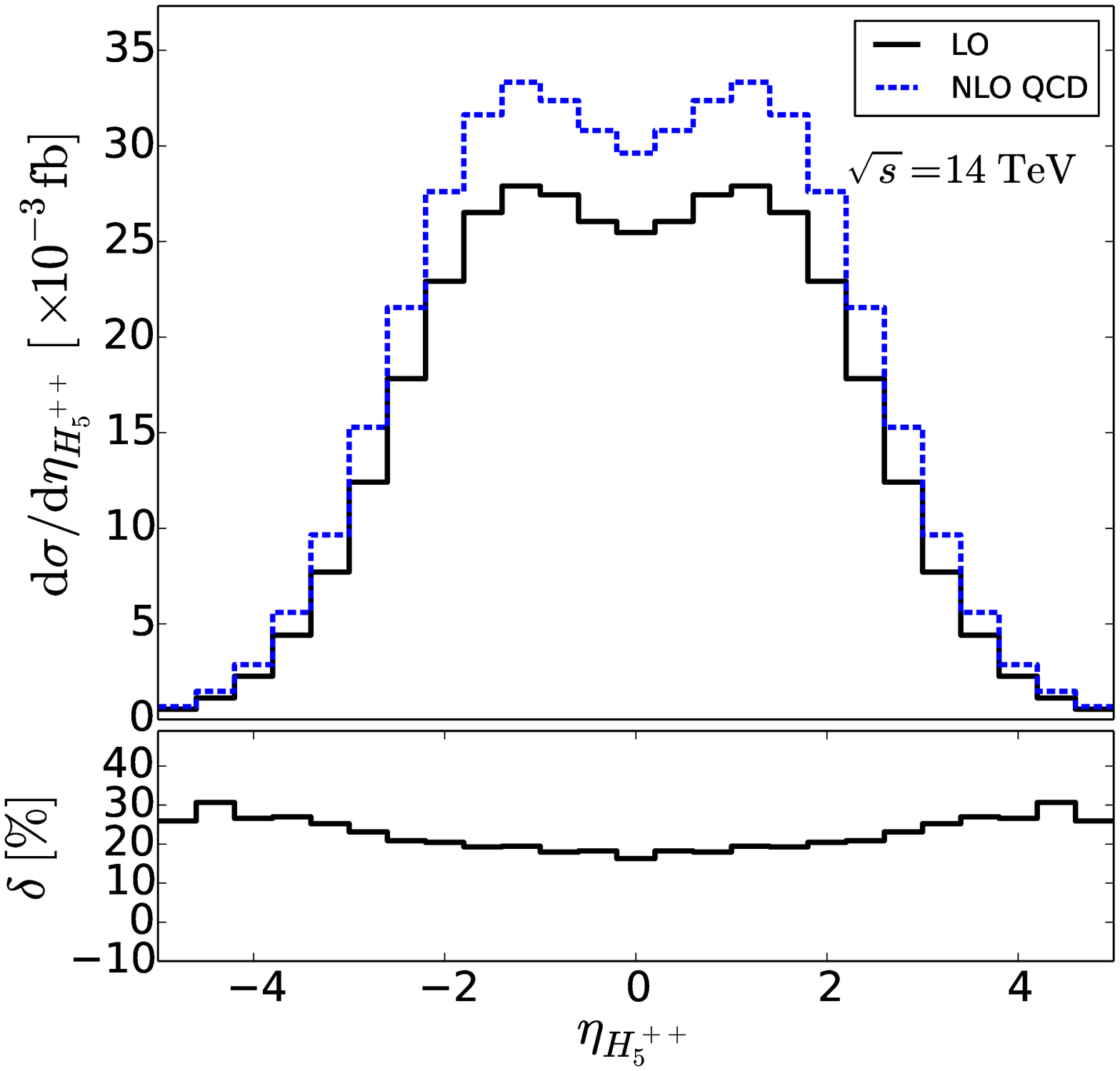}
\includegraphics[scale=0.35]{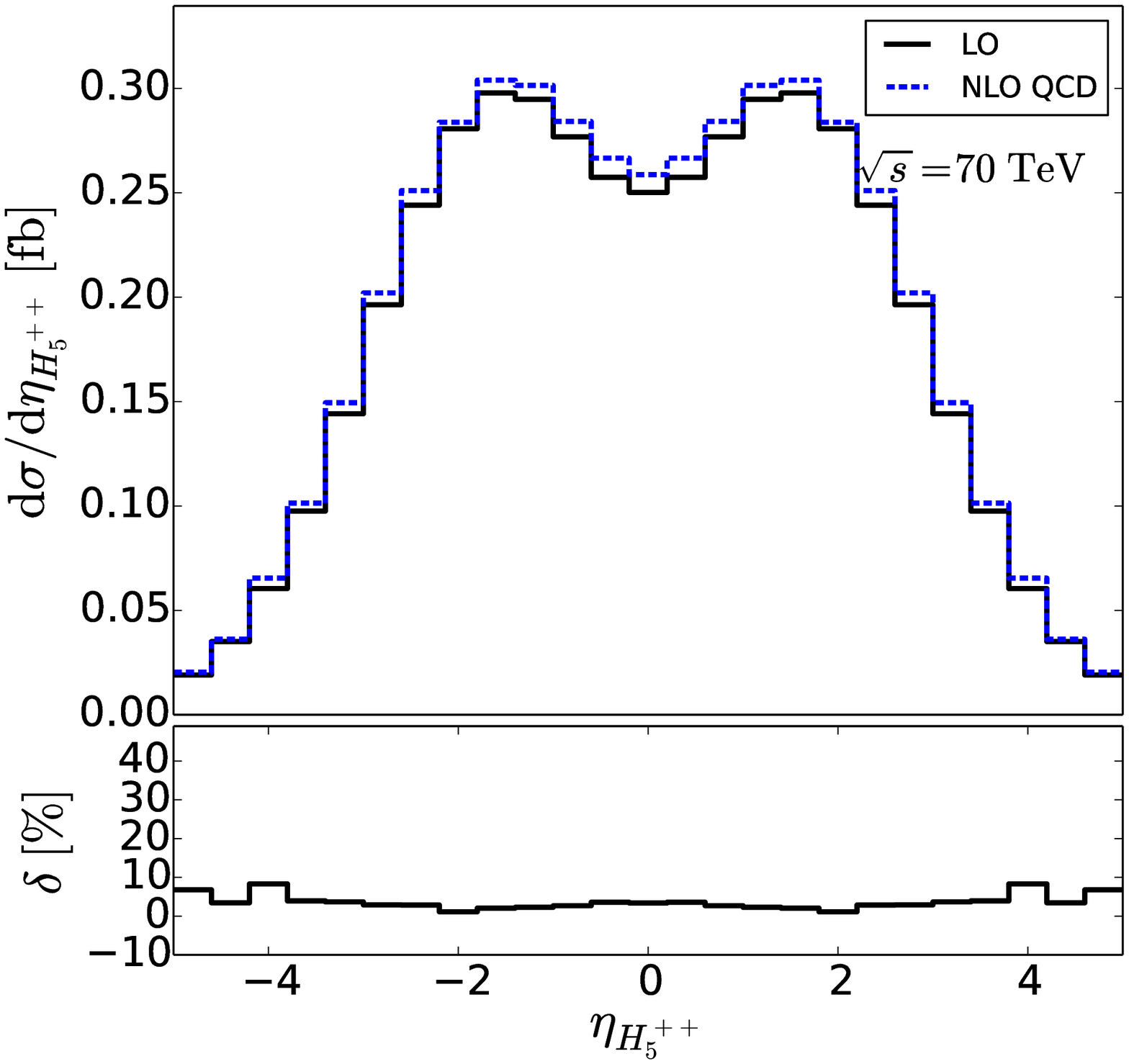}
\caption{Rapidity distributions of $H_5^{++}$ for $pp \rightarrow W^{+}W^{+} \rightarrow H_5^{++} h^0 + 2\, {\rm jets}$ at hadron colliders.}
\label{rapidity-H5}
\end{center}
\end{figure}

The LO and NLO QCD corrected pseudorapidity distributions of the leading jet, $h^0$, and $H_5^{++}$ and the corresponding QCD relative corrections are displayed in Figs.\ref{rapidity-jet}-\ref{rapidity-H5}, respectively. From the left plot of Fig.\ref{rapidity-jet} we can see that both the LO and NLO QCD corrected pseudorapidity distributions of the leading jet peak at $|\eta_{j_1}| \sim 3$ at the $14~ {\rm TeV}$ LHC. As expected, the leading jet prefers to be produced in the forward and backward regions for a VBF event at the LHC. At the $70~ {\rm TeV}$ SPPC, the final leading jet tends to be more collinear to the incoming protons compared to that at the $14~ {\rm TeV}$ LHC. As $|\eta_{j_1}|$ increases from $0$ to $4.4$, the QCD relative corrections increase from $-8\%$ to $40\%$ and from $-20\%$ to $17\%$ at the $14~ {\rm TeV}$ LHC and $70~ {\rm TeV}$ SPPC, respectively. Figures.\ref{rapidity-h0} and \ref{rapidity-H5} show that the final Higgs bosons ($h^0$ and $H_5^{++}$) are mostly produced in the central pseudorapidity region. At the $14~ {\rm TeV}$ LHC, the NLO QCD correction enhances the LO $\eta_{h^0}$ and $\eta_{H_5^{++}}$ distributions significantly and the QCD relative corrections increase from $16\%$ to $36\%$ and from $16\%$ to $31\%$ as $|\eta_{h^0}|$ and $|\eta_{H_5^{++}}|$ increase from $0$ to $5$, respectively. The pseudorapidity distributions of final Higgs bosons at the $70~ {\rm TeV}$ SPPC are similar to those at the $14~ {\rm TeV}$ LHC, but the QCD relative corrections are relatively small and are steady at about $3\%$.

\par
\begin{figure}[htbp]
\begin{center}
\includegraphics[scale=0.35]{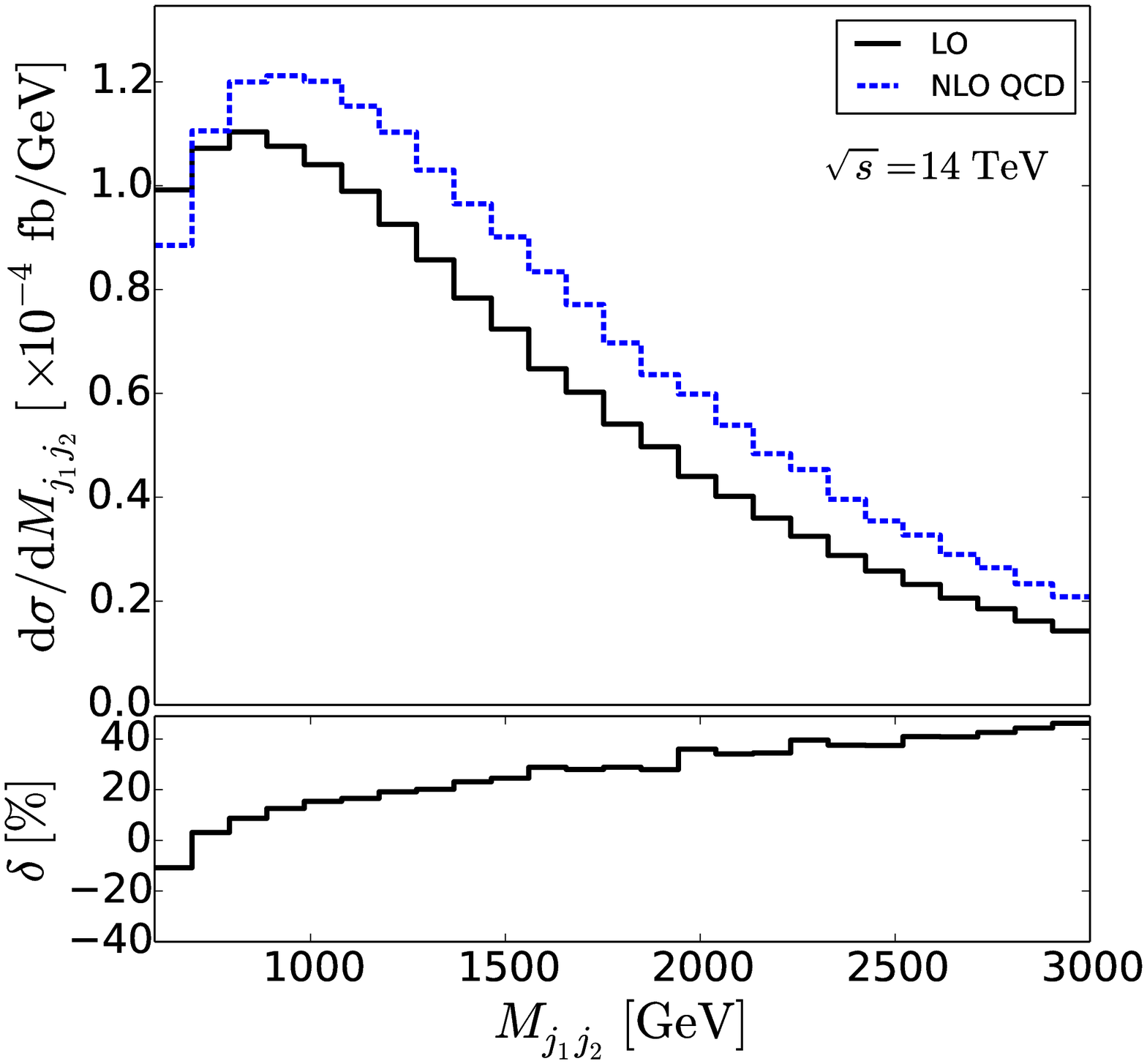}
\includegraphics[scale=0.35]{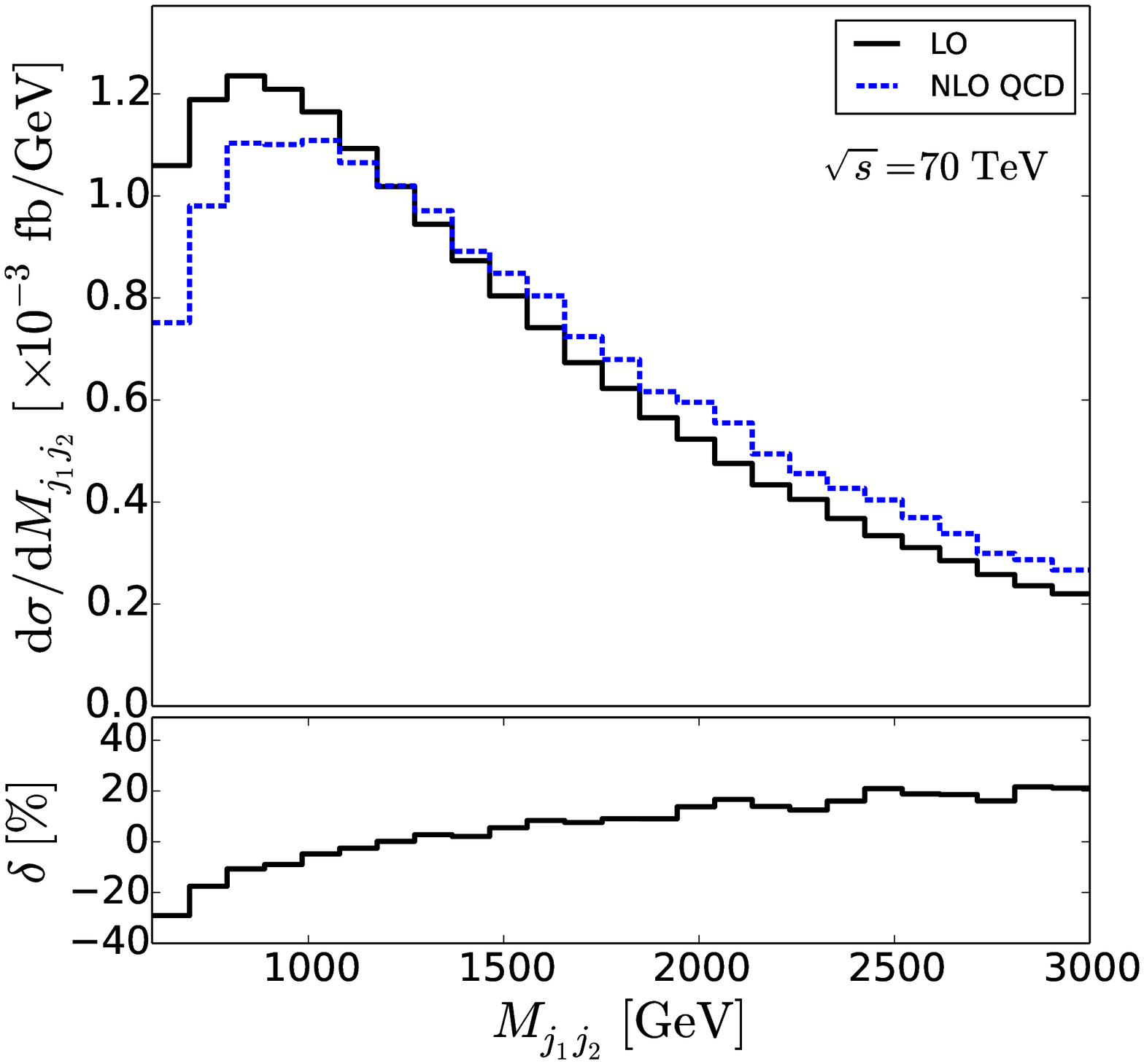}
\caption{Invariant mass distributions of the two hardest final-state jets for $pp \rightarrow W^{+}W^{+} \rightarrow H_5^{++} h^0 + 2\, {\rm jets}$ at hadron colliders.}
\label{invmass}
\end{center}
\end{figure}

The LO and NLO QCD corrected invariant mass distributions of the two hardest final-state jets and the corresponding QCD relative corrections are plotted in Fig.\ref{invmass}. At both LO and QCD NLO, the invariant mass distributions increase rapidly at first and then decrease slowly as the increment of $M_{j_1j_2}$. Although $K > 1$ for the integrated cross section, the NLO QCD correction might suppress the LO differential cross section in some kinematic region. The right bottom panel of Fig.\ref{invmass} shows that the QCD relative correction to the invariant mass distribution at the $70~ {\rm TeV}$ SPPC is negative in the region of $M_{j_1j_2} < 1.2~ {\rm TeV}$. At the $14~ {\rm TeV}$ LHC and $70~ {\rm TeV}$ SPPC, the QCD relative corrections increase from $-12\%$ to $48\%$ and from $-29\%$ to $20\%$, respectively, as $M_{j_1j_2}$ increases from $600~ {\rm GeV}$ to $3~ {\rm TeV}$.

\subsection{Signal and background}
For the VBF production of $H_5^{\pm\pm} h^0$ at $pp$ colliders considered in this paper, the signal process is chosen as\footnote{$\ell_1$ and $\ell_2$ are called leading and next-to-leading leptons, respectively, according to their transverse momentum in decreasing order, i.e., $p_{T,\ell_1} > p_{T,\ell_2}$.}
\begin{eqnarray}
pp \rightarrow W^{\pm}W^{\pm} \rightarrow H_5^{\pm\pm} \, \Big[ \rightarrow W^{\pm}W^{\pm} \rightarrow \ell^{\pm}_1 \ell^{\pm}_2 \overset{ _{(-)}}{\nu_{\ell_1}} \overset{ _{(-)}}{\nu_{\ell_2}} \, \Big] \, h^0 + 2\, {\rm jets}~~~~~(\ell_1, \ell_2 = e ~ {\rm or}~ \mu).
\end{eqnarray}
Because of the smallness of the lepton Yukawa coupling $H_5^{\pm\pm} \ell^{\mp} \ell^{\mp}$, the branching ratio for $H_5^{\pm\pm} \rightarrow W^{\pm}W^{\pm}$ is almost $100\%$ if this decay mode is kinematically allowed.
The vector-boson-associated (VBA) production of $H_5^{\pm\pm} h^0$, i.e., $pp \rightarrow H_5^{\pm\pm} h^0 W^{\mp} + X$, can also provide the same final state as the signal process via the following cascade decays,
\begin{eqnarray}
H_5^{\pm\pm} \rightarrow W^{\pm}W^{\pm} \rightarrow \ell^{\pm}_1 \ell^{\pm}_2 \overset{ _{(-)}}{\nu_{\ell_1}} \overset{ _{(-)}}{\nu_{\ell_2}}
~~~~~~~{\rm and}~~~~~~~
W^{\mp} \rightarrow 2\, {\rm jets},
\end{eqnarray}
but is not taken into account in the signal-background analysis because
\begin{eqnarray}
\frac{\sigma\Big(pp \rightarrow H_5^{\pm\pm} h^0 W^{\mp}\, \big[\rightarrow 2\, {\rm jets} \, \big]\Big)}{\sigma\Big(pp \rightarrow W^{\pm}W^{\pm} \rightarrow H_5^{\pm\pm} h^0 + 2\, {\rm jets}\Big)}
< 0.1 \%
\end{eqnarray}
after applying the VBF cuts. Therefore, the background to the VBF production of $H_5^{\pm\pm} h^0$ mainly comes from the SM process $pp \rightarrow W^{\pm} W^{\pm} h^0 + 2\, {\rm jets} + X$ with subsequent leptonic decays of $W$ bosons.

\par
\begin{figure}[htbp]
\begin{center}
\includegraphics[scale=0.31]{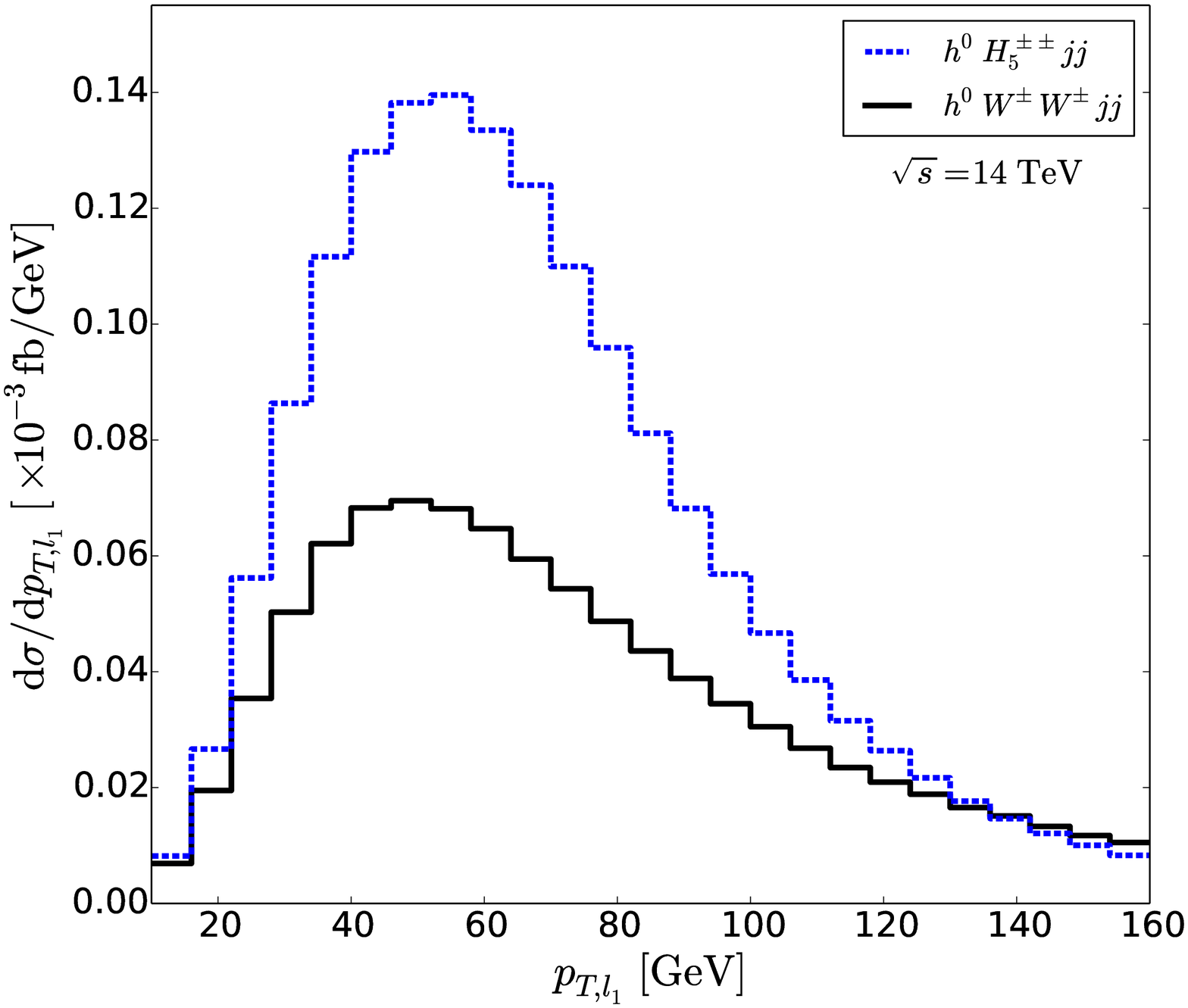}
\includegraphics[scale=0.31]{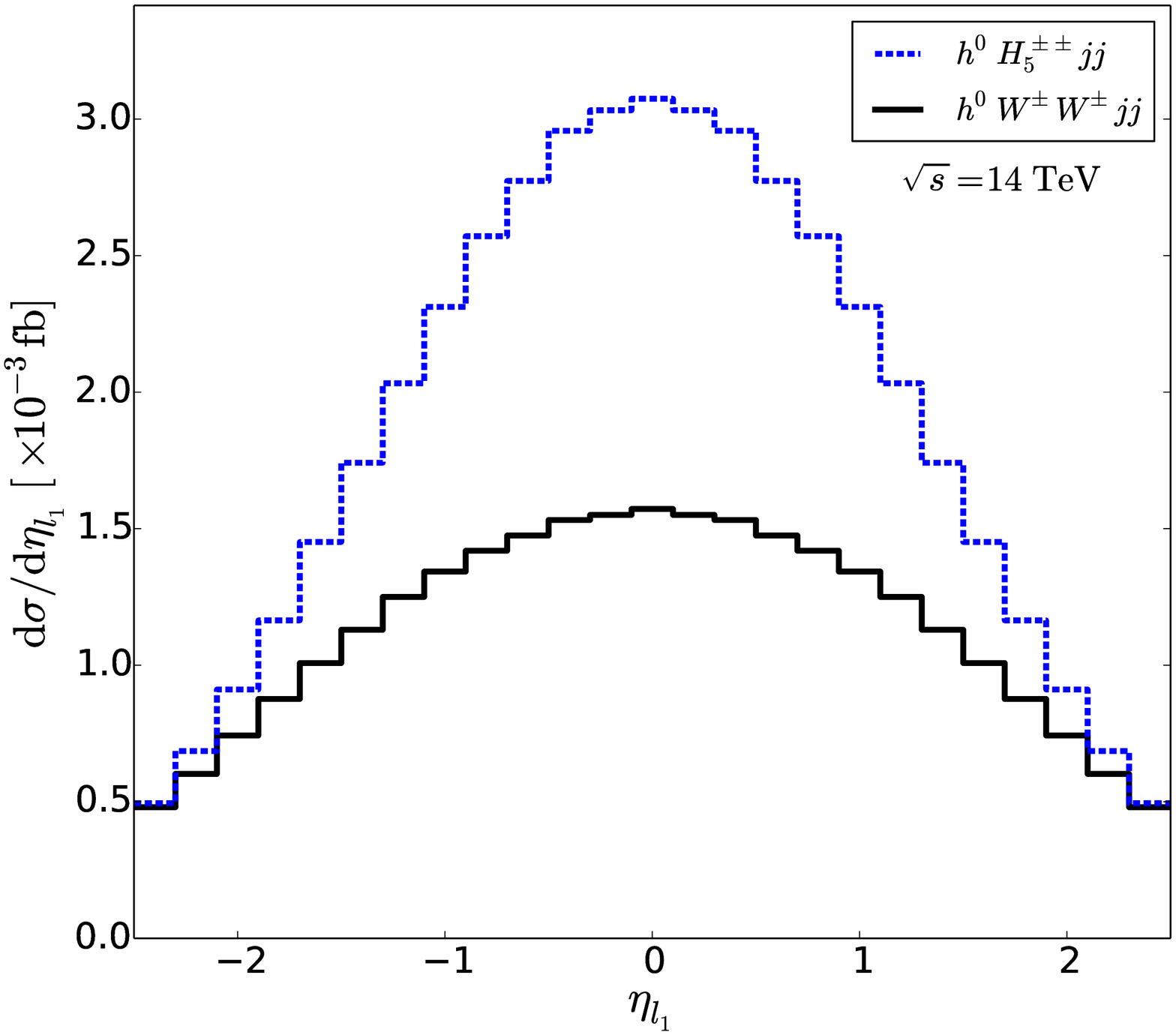}
\includegraphics[scale=0.31]{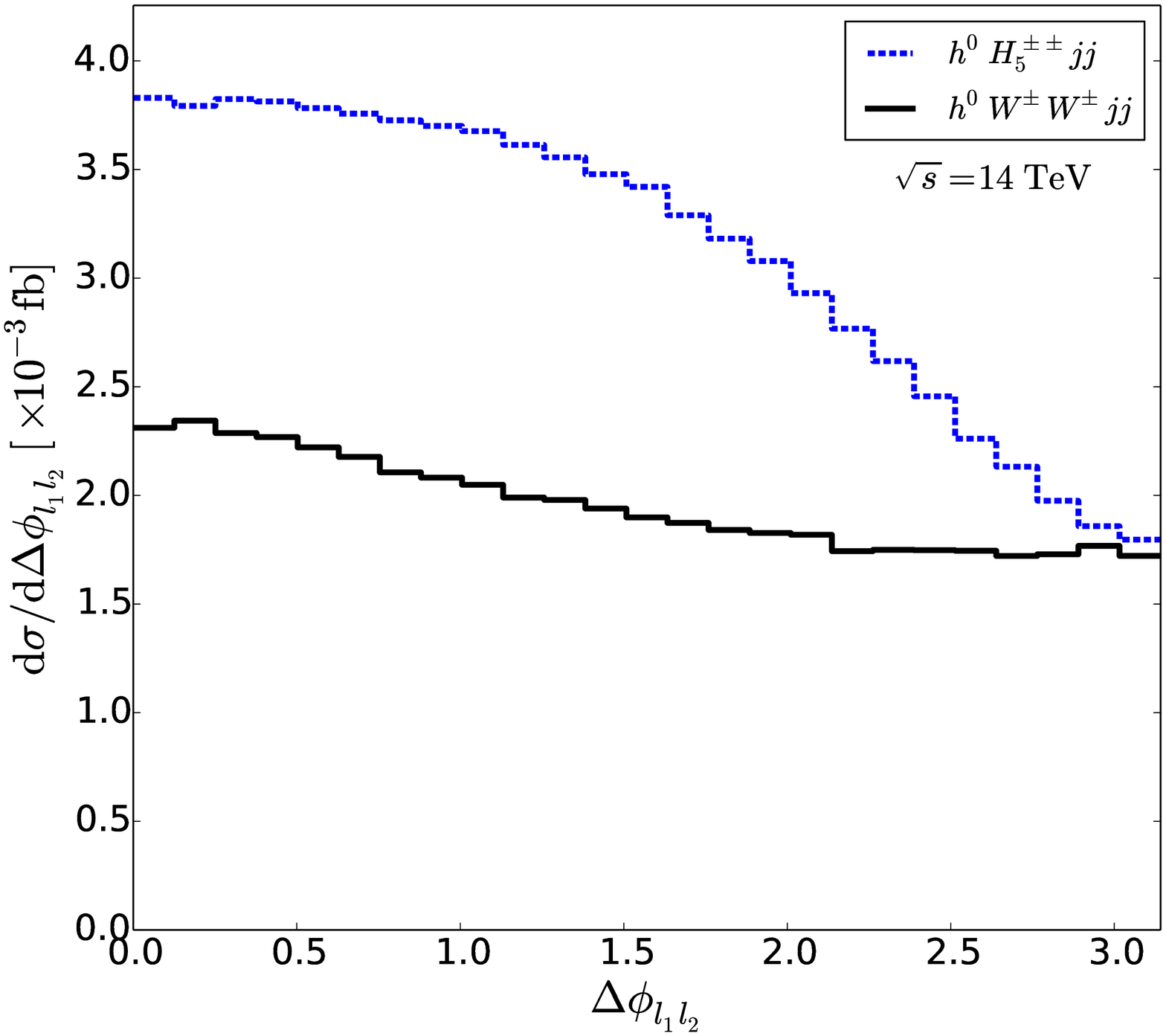}
\includegraphics[scale=0.31]{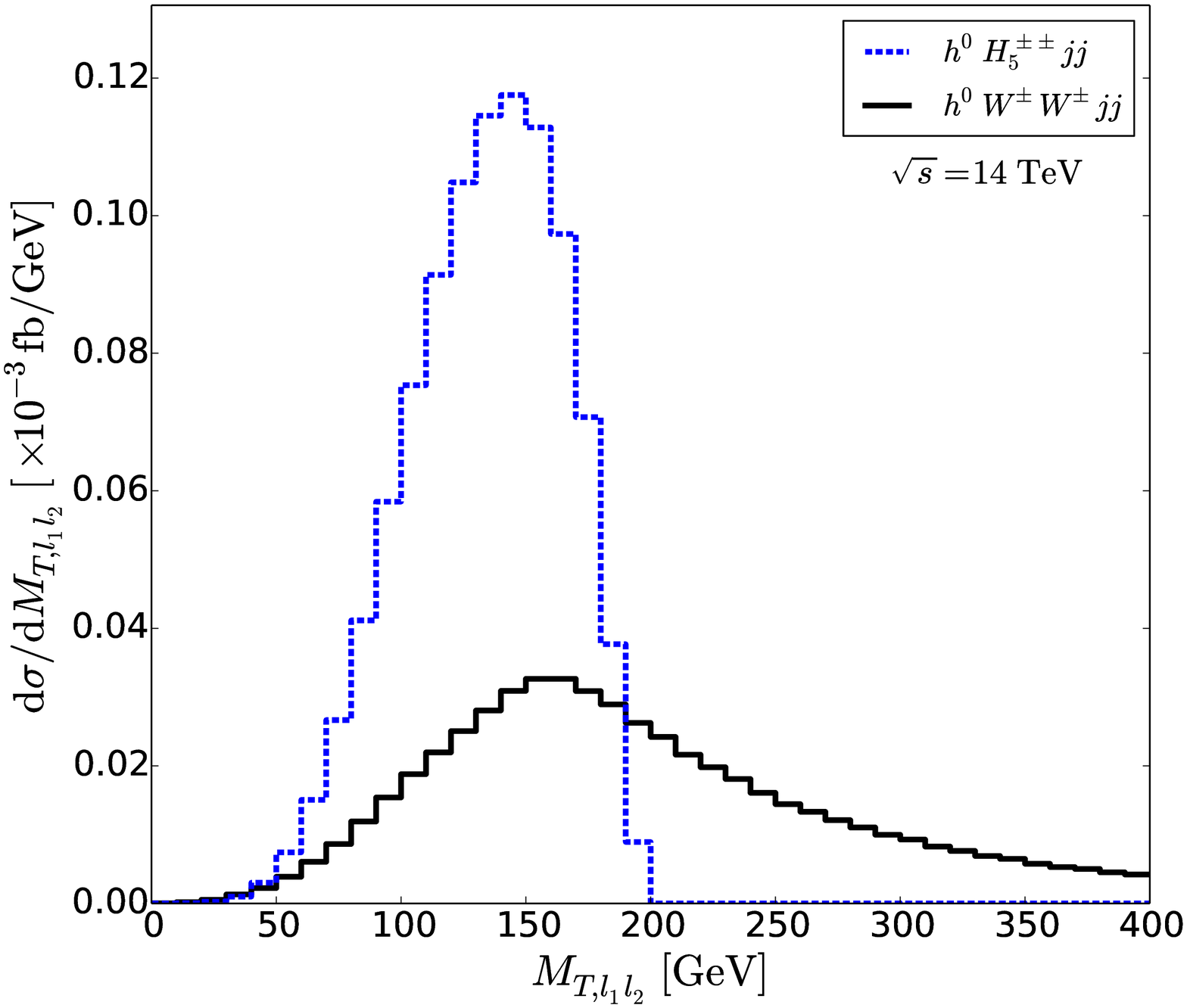}
\caption{Differential distributions of the final leptons for both signal and SM background at the $14~ {\rm TeV}$ LHC.}
\label{dis-SB}
\end{center}
\end{figure}

The signal and SM background events are generated by using the {\sc MadGraph5} package, and the spin correlation and finite width effects of the intermediate Higgs and $W$ bosons are taken into account by employing the {\sc madspin} \cite{madspin1,madspin2} method. When calculating the cross sections and generating the event samples for both signal and background, the kinematic and geometric acceptance requirements of
\begin{eqnarray}
p_{T,\ell} > 10~ {\rm GeV},~~~~~~~ |\eta_{\ell}| < 2.5
\end{eqnarray}
on the final leptons and the VBF cuts are imposed as baseline cuts. In Fig.\ref{dis-SB}, we present the $p_{T,\ell_1}$, $\eta_{\ell_1}$, $\Delta \phi_{\ell_1\ell_2}$, and $M_{T, \ell_1\ell_2}$ distributions for both signal and SM background at the $14~ {\rm TeV}$ LHC, respectively. $p_{T,\ell_1}$ and $\eta_{\ell_1}$ are the transverse momentum and pseudorapidity of the leading lepton $\ell_1$, $\Delta \phi_{\ell_1\ell_2}$ is the azimuthal separation of the two final-state same-sign leptons, and $M_{T, \ell_1\ell_2}$ is the transverse mass given by \cite{ww-interacting,heavy-leptons}
\begin{eqnarray}
M_{T, \ell_1 \ell_2}
=
\sqrt{
\left[ \sqrt{M_{\ell_1 \ell_2}^2 + p_{T, \ell_1 \ell_2}^2} + \slashed{p}_T \right]^2 - \left| \vec{p}_{T, \ell_1 \ell_2} + \vec{\slashed{p}}_T \right|^2
} \,\, .
\end{eqnarray}
These kinematic distributions show that the line shapes of $p_{T,\ell_1}$, $\eta_{\ell_1}$, and $\Delta \phi_{\ell_1\ell_2}$ distributions of the signal are similar to those of the SM background, and the integrated cross section of the signal process is about twice as large as that of the SM background at the $14~ {\rm TeV}$ LHC. The differential cross section of the signal peaks at $M_{T, \ell_1 \ell_2} \sim 140~ {\rm GeV}$ and decreases sharply to zero as $M_{T, \ell_1 \ell_2}$ increases from $140~ {\rm GeV}$ to about $200~ {\rm GeV}$. After applying the transverse mass cut of
\begin{eqnarray}
\label{TM-cut}
M_{T, \ell_1 \ell_2} < 190~ {\rm GeV},
\end{eqnarray}
the SM background can be suppressed significantly, while almost all the signal events can pass this transverse mass cut.

\par
The significance of signal over background ${\cal S}$ is defined as
\begin{align}
{\cal S} = \frac{N_S}{\sqrt{N_S + N_B}}
= \frac{\sigma_S}{\sqrt{\sigma_S + \sigma_B}} \sqrt{{\cal L}} \,,
\end{align}
where $N_{S,B}$ and $\sigma_{S,B}$ are the event numbers and the cross sections for signal and background, respectively, and ${\cal L}$ represents the integrated luminosity. In Table \ref{signal-background} we provide the cross sections for the signal and the SM background before and after applying the event selection criterion in Eq.(\ref{TM-cut}), as well as the significance based on ${\cal L} = 100$ and $3000~{\rm fb}^{-1}$ which can be accumulated at the future High-Luminosity Large Hadron Collider \cite{hl-lhc}. The kinematic cuts I and II in Table \ref{signal-background} represent ``baseline cuts" and ``baseline cuts + event selection criterion (transverse mass cut)," respectively. This table shows that the significance is only about $0.8$ with $100~ {\rm fb}^{-1}$ integrated luminosity and can exceed four with $3000~ {\rm fb}^{-1}$ integrated luminosity at the $14~ {\rm TeV}$ $pp$ collider by imposing only the baseline cuts. As the colliding energy increases to $70~ {\rm TeV}$, the cross sections for the signal and the SM background increase to about $0.15$ and $0.2~ {\rm fb}$, and the significance can reach about $2.5$ and $13.6$ with ${\cal L} = 100$ and $3000~{\rm fb}^{-1}$, respectively. After applying the transverse mass cut on the two final-state same-sign leptons, the SM background is reduced over $50\%$, but the loss of signal is less than $1\%$, especially at very high-energy $pp$ colliders. This large suppression of the SM background by the event selection criterion improves the significance of $H_5^{\pm\pm} h^0$ VBF signal. For example, the significance can reach about $3.1$ by applying the event selection criterion at the $70~ {\rm TeV}$ SPPC with $100~ {\rm fb}^{-1}$ integrated luminosity. At the $14~ {\rm TeV}$ LHC and $70~ {\rm TeV}$ SPPC, the luminosities required for the ${\cal S} \geq 5$ discovery of the $H_5^{\pm\pm} h^0$ VBF signal are about $4100$ and $400~ {\rm fb}^{-1}$ by imposing only the baseline cuts, and can be reduced to about $3400$ and $250~ {\rm fb}^{-1}$ by further applying the transverse mass cut. It is concluded that the $H_5^{\pm\pm} h^0$ VBF signal can be directly detected at future high-luminosity, high-energy hadron colliders if the dynamics of beyond the SM physics is governed by the GM model.
\begin{table}[htbp]
\begin{center}
\renewcommand\arraystretch{1.5}
\begin{tabular}{|c|c|c|c|c|c|c|}
\hline
\multirow{2}*{~Cuts~}  &  \multicolumn{3}{c}{$\sqrt{S} = 14~ {\rm TeV}$}  &  \multicolumn{3}{|c|}{$\sqrt{S} = 70~ {\rm TeV}$}  \\
\cline{2-7}
  &  $\sigma_S$ (fb)  &  $\sigma_B$ (fb)  &  ${\cal S}$  &  $\sigma_S$ (fb)  &  $\sigma_B$ (fb)  &  ${\cal S}$  \\
\hline
\multirow{2}*{I}   &  \multirow{2}*{$0.00984$}  &  \multirow{2}*{$0.00615$}  &  $~0.778~~~ \scriptstyle{(100~ {\rm fb}^{-1})}~~\,$
                   &  \multirow{2}*{$0.147$}   &  \multirow{2}*{$0.203$}    &  $~2.48~~~ \scriptstyle{(100~ {\rm fb}^{-1})}~~\,$  \\
                                                                      &  &  &  $4.26~~~~ \scriptstyle{(3000~ {\rm fb}^{-1})}$
                                                                      &  &  &  $13.6~~~ \scriptstyle{(3000~ {\rm fb}^{-1})}$  \\
\hline
\multirow{2}*{II}  &  \multirow{2}*{$0.00975$}  &  \multirow{2}*{$0.00300$}  &  $0.863~~~ \scriptstyle{(100~ {\rm fb}^{-1})}~\,$
                   &  \multirow{2}*{$0.145$}   &  \multirow{2}*{$0.0698$}   &  $3.13~~~ \scriptstyle{(100~ {\rm fb}^{-1})}~\,$  \\
                                                                      &  &  &  $4.73~~~~ \scriptstyle{(3000~ {\rm fb}^{-1})}$
                                                                      &  &  &  $17.1~~~ \scriptstyle{(3000~ {\rm fb}^{-1})}$  \\
\hline
\end{tabular}
\caption{Cross sections for signal and SM background as well as the significance based on ${\cal L} = 100$ and $3000~ {\rm fb}^{-1}$ before (I) and after (II) applying the event selection criterion in Eq.(\ref{TM-cut}) at the $14$ and $70~ {\rm TeV}$ $pp$ colliders.}
\label{signal-background}
\end{center}
\end{table}

\par
\section{SUMMARY}
The existence of a doubly charged Higgs boson is a distinct feature of the GM model. In this work, we perform a parameter scan of the GM model on the $m_5-v_{\Delta}$ plane and investigate in detail the doubly charged Higgs boson production in association with a SM-like Higgs boson via VBF in the H5plane benchmark scenario at the $14~ {\rm TeV}$ LHC and $70~ {\rm TeV}$ SPPC. Both the integrated cross section and the distributions with respect to some kinematic variables of final Higgs bosons and leading jet are provided up to the QCD NLO. The numerical results show that the NLO QCD correction can enhance the total cross section for $pp \rightarrow W^{+}W^{+} \rightarrow H_5^{++} h^0 + 2\, {\rm jets}$ by about $20\%$ and $3 \sim 5\%$ at the $14~ {\rm TeV}$ LHC and $70~ {\rm TeV}$ SPPC, respectively. The theoretical uncertainty due to the renormalization, factorization scale is underestimated at the LO at the $70~ {\rm TeV}$ SPPC, while it can be suppressed significantly by the NLO QCD correction at the $14~ {\rm TeV}$ LHC. In the signal-background analysis, we adopt the {\sc madspin} method to take into account the spin correlation and finite width effects in dealing with the cascade decay of the doubly charged Higgs boson $H_5^{\pm\pm} \rightarrow W^{\pm}W^{\pm} \rightarrow \ell^{\pm}_1 \ell^{\pm}_2 \overset{ _{(-)}}{\nu_{\ell_1}} \overset{ _{(-)}}{\nu_{\ell_2}}$ and provide the distributions of final leptons for the signal and the SM background. We find that the SM background can be reduced remarkably and the significance of the $H_5^{\pm\pm} h^0$ VBF process can be improved by imposing a proper cut on the transverse mass $M_{T, \ell_1 \ell_2}$. This $H_5^{\pm\pm} h^0$ VBF signal can be directly detected at future high-luminosity, high-energy hadron colliders if the Higgs sector of new physics is described by the GM model.

\vskip 5mm
\par
\noindent{\large\bf Acknowledgments} \\
This work is supported in part by the National Natural Science Foundation of China (Grants No. 11775211, No. 11535002, No. 11375171, and No. 11675033), the Fundamental Research Funds for the Central Universities (Grant No. DUT18LK27), and the CAS Center for Excellence in Particle Physics (CCEPP).

\end{document}